\documentclass[aps,pra,nofootinbib,reprint,preprintnumbers,superscriptaddress,longbibliography,floatfix]{revtex4-2}

\usepackage{latexsym,epsfig,amssymb,amsfonts,amsmath,graphicx,bbm,braket,esdiff,dsfont,array,hyperref}

\usepackage[export]{adjustbox}
\pdfoutput=1

\hypersetup{
pdftitle={Classical verification of a quantum simulator}, pdfsubject={Local relaxation of a Bose gas}, pdfauthor={Paul Secular}, pdfkeywords={verification, replication, validation, computational, physics, relaxation, thermalization, strongly-correlated, quantum, tensor network, tensor train, Hamiltonian, time evolution, simulation, DMRG, t-DMRG, TDMRG, TEBD, MPS, MPI, OpenMP}, colorlinks=true, allcolors=blue}

\usepackage{booktabs}
\usepackage{xcolor}
\usepackage[percent]{overpic}
\usepackage[shortlabels]{enumitem}
\usepackage[english]{babel}
\usepackage{afterpage}
\usepackage[subrefformat=parens,labelformat=parens,caption=false]{subfig}

\renewcommand{\selectlanguage}[1]{}

\addto\captionsenglish{

}

\usepackage{soul}

\newcommand{\subfir}[1]{Fig.\@~\subref*{#1}}
\newcommand{\subfirs}[1]{Figs.\@~\subref*{#1}}
\newcommand{\subtab}[1]{Table~\subref*{#1}}

\makeatletter
\newcommand{\colorcaption}[2][]{
  \begingroup
  \renewcommand{\@caption@fignum@sep}{ (color online). }
  \caption[#1]{#2}
  \endgroup
}
\makeatother

\allowdisplaybreaks

\usepackage{environ}
\usepackage{tikz}
\usetikzlibrary{calc,matrix}

\newcounter{lines}

\newcounter{vtml}
\begin{document}

\title{Classical verification of a quantum simulator: local relaxation of a 1D Bose gas}

\author{Paul Secular}
\email{paul@secular.me.uk}
\homepage{http://secular.me.uk/}
\affiliation{Department of Mathematical Sciences, University of Bath, Claverton Down, Bath BA2 7AY, UK}

\date{\today}

\begin{abstract}
In [Nat. Phys. 8, 325–330 (2012)], Trotzky \emph{et al.\@} utilize ultracold atoms in an optical lattice to simulate the local relaxation dynamics of a strongly interacting Bose gas ``for longer times than present classical algorithms can keep track of''. Here, I classically verify the results of this analog quantum simulator by calculating the evolution of the same quasi-local observables up to the time at which they appear ``fully relaxed''. Using a parallel implementation of the time-evolving block decimation (TEBD) algorithm to simulate the system on a supercomputer, I show that local densities and currents can be calculated in a matter of days rather than weeks. The precision of these numerics allows me to observe deviations from the conjectured power-law decay and to determine the effects of the harmonic trapping potential. As well as providing a robust benchmark for future experimental, theoretical, and numerical methods, this work serves as an example of the independent verification process.
\end{abstract}

\maketitle

\section{Introduction}\newcounter{fnnumber}

In the quest for practical quantum advantage~\cite{daley_practical_2022, hoefler_disentangling_2023}, the verification of noisy quantum simulators remains an important task~\cite{shaffer_practical_2021}. In this work, I revisit a landmark paper~\cite{trotzky_probing_2012} that describes ``the first dynamical quantum simulator''~\cite{noauthor_theoretical_nodate}. This remarkable experiment used ultracold atoms in an optical lattice to simulate the dynamics of a one-dimensional (1D) strongly interacting Bose gas. Here, I ask to what extent these results can be verified classically on a supercomputer using quasi-exact tensor network methods. Employing Vidal's parallel time-evolving block decimation (pTEBD) algorithm~\cite{vidal_efficient_2003, vidal_efficient_2004, urbanek_parallel_2016, sun_improved_2023}, I extend the matrix product state (MPS) simulations of Ref.~\cite{trotzky_probing_2012} to later times, finding that it is possible to accurately compute expectation values up to the point at which they appear relaxed experimentally. In contrast to the paper's numerics, which took about five weeks of runtime at the J{\"u}lich Supercomputing Centre~\cite{eisert_towards_2017}, each of my simulations required less than 50 hours of runtime on the University of Bath's now retired Balena supercomputer~\cite{university_of_bath_balena_nodate}. This highlights the importance of using parallel algorithms to take advantage of classical high-performance computing (HPC) clusters~\cite{secular_parallel_2024}.

\subsection{Model}

Starting from a charge density wave (CDW) product state, the experiment described in Ref.~\cite{trotzky_probing_2012} (henceforth, ``the experiment'') aims to simulate the local relaxation dynamics of the 1D Bose--Hubbard model~\cite{jaksch_cold_1998} in the presence of a harmonic trap. However, as the tight-binding approximation is found to break down for small $U/J$, the authors add a next-nearest-neighbor (NNN) hopping term to the Hamiltonian $\hat{H}$. The model actually being simulated is thus given by
\begin{eqnarray}\label{eqn:trotzky_Hamiltonian}
    \hat{H} &=& \frac{U}{2} \sum_{j=1}^{L}{\hat{n}_j (\hat{n}_j - 1)}
    -J \sum_{j=1}^{L-1} (\hat{b}_j^\dagger \hat{b}_{j+1} + \hat{b}_{j+1}^\dagger \hat{b}_{j})\\* \nonumber
    &+& \frac{K}{2} \sum_{j=1}^L{(j-j_c)^2 \hat{n}_j}
    - J_{\mathrm{NNN}} \sum_{j=1}^{L-2} (\hat{b}_j^\dagger \hat{b}_{j+2} + \hat{b}_{j+2}^\dagger \hat{b}_{j}),
\end{eqnarray}
where $J_{\mathrm{NNN}}$ is the NNN hopping amplitude, $K$ is the strength of the trap, and $j_c$ is the central lattice site. The experimental time is given in units of $h/4J = \pi \hbar/2J$,\footnote{Contrast this with Ref.~\cite{schmitteckert_relaxation_2012}, where the unit of time is smaller by a factor of $\pi/2$.} which is approximately equal to $1\,\mathrm{ms}$ in SI units (i.e.\@ $J \approx \pi \hbar/2\,\mathrm{ms}^{-1}$)~\cite{trotzky_probing_2012}. Setting $J = \hbar = 1$ here means I measure time in units of $\pi/2$.

\subsection{Relation to previous work}

The analytic solution of the non-interacting nearest-neighbor (NN) version of this problem was presented in 2008~\cite{cramer_exact_2008}. A numerical study of the interacting NN problem followed shortly afterwards~\cite{cramer_exploring_2008, flesch_probing_2008}. This no doubt motivated the experimental paper~\cite{trotzky_probing_2012}, in which the authors also provide NN and NNN numerics. A swift response~\cite{schmitteckert_relaxation_2012} aimed ``to demonstrate that simulations on classical computers based on matrix product states can be performed reliably on a time scale which exceeds the time scale of the reported experimental data''. However, the simulations of Ref.~\cite{schmitteckert_relaxation_2012} do not exceed the experimental timescale, or indeed the numerical timescale of Refs.~\cite{cramer_exploring_2008, flesch_probing_2008}.

The present work was inspired by a 2016 benchmark calculation~\cite[Fig.~7]{urbanek_parallel_2016}, which showed that the experimental relaxation timescale can in fact be reached classically using pTEBD---at least in the NN case. However, this experiment is not the main focus of Ref.~\cite{urbanek_parallel_2016}, so its authors only calculate a single observable for a single set of parameters with no NNN hopping. The results I present here reach the same timescale while taking account of NNN hopping. They are thus the first to allow direct verification of the experimental quantum simulator results of Ref.~\cite[Figs.~2,~3(c),~4(b)]{trotzky_probing_2012} beyond the short times reached numerically in the original paper.

In a sign of continued interest in this problem, two more recent works directly compare the experimental data of Ref.~\cite{trotzky_probing_2012} to analytic calculations. Firstly, in a 2017 report, Moussa \emph{et al.\@} use the time-dependent variational principle to analytically derive a mean-field Hamiltonian, ignoring NNN hopping but including the trapping potential~\cite[Fig.~5.1]{moussa_realizing_2017}. Secondly, in a 2020 paper, Dabelow and Reimann apply a novel perturbative relaxation theory with a single fitting parameter to a number of systems, including the NN Bose--Hubbard model~\cite[Fig.~2]{dabelow_relaxation_2020}. However, as their theory is formulated in the thermodynamic limit, it ignores the harmonic trapping potential (as well as the NNN hopping)~\cite{secular_comment_2020}. A benefit of the present work is that it allows for a quantitative assessment of the accuracy of these approaches, as well as providing a benchmark for future analytic techniques.

\section{Method}

\subsection{Experimental details}

\begin{table}[!b]
    \centering
    \subfloat[\label{stab:trotzky_params}]{
    \begin{tabular}{ccccc}
        \toprule
          Experiment & $U/J$ & $J_{\text{NNN}}/J$ & $K/J$ & (preprint) \\ 
        \midrule
          a & $2.44(2)$ & $0.12$ & $0.01$ & $0.005$ \\
          b & $3.60(4)$ & $0.08$ & $0.013$ & $0.007$ \\
          c & $5.16(7)$ & $0.05$ & $0.017$ & $0.009$ \\
          d & $9.9(1)$ & $0.03$ & $0.029$ & $0.015$ \\
        \bottomrule
    \end{tabular}}\\
    \subfloat[\label{stab:secular_params}]{
    \begin{tabular}{ccccc}
        \toprule
          pTEBD simulation & $U$ & $J_{\text{NNN}}$ & $K$ & $J$ \\
        \midrule
          a & $2.442$ & $-0.12$ & $0.010$ & 1\\
          b & $3.604$ & $-0.08$ & $0.013$ & 1\\
          c & $5.167$ & $-0.05$ & $0.017$ & 1\\
          d & $9.910$ & $-0.03$ & $0.029$ & 1\\
        \bottomrule
    \end{tabular}}
    \caption{\protect\subref{stab:trotzky_params} Experimental Hamiltonian parameters reported by Trotzky \emph{et al.\@} in Ref.~\cite{trotzky_probing_2012} for simulations [a, b, c, d]. The ``(preprint)'' column refers to the values of $K/J$ appearing in Refs.~\cite{trotzky_probing_2011, schmitteckert_relaxation_2012}. \protect\subref{stab:secular_params} Corresponding Hamiltonian parameters used in this work to verify the quantum simulations numerically.}
    \label{table:bh_params}
\end{table}

Rather than simulating a single chain, the experiment involves an ensemble of chains, each having a different total particle number $N$. The expectation values are thus ensemble averages. To verify their experimental results, the authors carry out $t$-DMRG~\cite{daley_time-dependent_2004, white_real-time_2004} calculations, reaching times of $t < 3\pi/2$. In contrast, they demonstrate that a timescale of ${t = 20\pi/2}$ (equivalent to 20 ms) can be reached with their quantum simulator~\cite[Fig.~1]{trotzky_probing_2012}. In the remainder of the paper, however, results are presented up to a time of $t \leq 5\pi/2$, since this is the point at which the measured observables appear relaxed to experimental precision. No finite-size recurrences~\cite{flesch_probing_2008} are observed after this time, which the authors attribute to the ensemble averaging and lack of sharp reflections~\cite{trotzky_probing_2012}.

Although Ref.~\cite[Figs.~3(c), 4(c)]{trotzky_probing_2012} provides summary results for a range of interaction strengths, the paper focuses on four sets of parameters in the intermediate $2 < U/J < 10$ regime. These are shown in \subtab{stab:trotzky_params}, along with the differing values of $K/J$ given in Ref.~\cite{trotzky_probing_2011}. In \subtab{stab:secular_params}, I list the values used in this work to verify the quantum simulations. Note the opposite signs for $J_{\text{NNN}}$ (as discussed in Section~\ref{sec:nnn-hopping}).

\begin{figure}[!b]
    \centering
    \includegraphics[width=8.5cm]{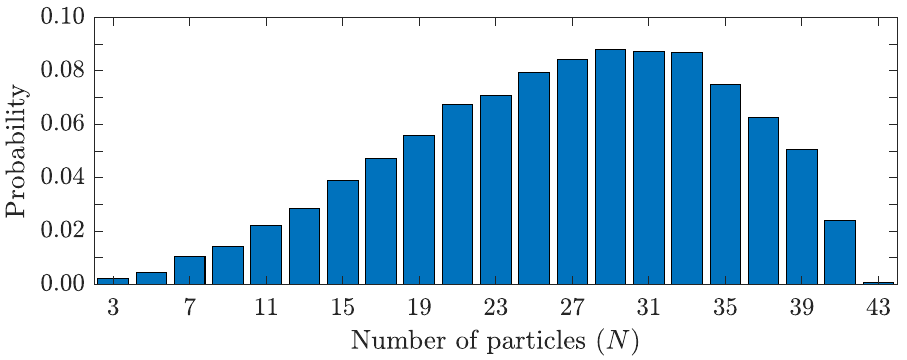}
    \colorcaption{\label{fig:trotzky-probabilities}Probability distribution for the total particle number $N$ in the experiment. The values are from Ref.~\cite[Supp.~Fig.~1]{trotzky_probing_2012}.}
\end{figure}

\subsection{Calculation details}

I follow Ref.~\cite{trotzky_probing_2012} in modeling the experiment by an ensemble of independent 121-site chains, each having an odd number of bosons $3 \leq N \leq 43$. I take the relative frequencies for $N$ from Ref.~\cite[Supp.~Fig.~1]{trotzky_probing_2012} (replotted in Fig.~\ref{fig:trotzky-probabilities}), finding a mean particle number of $\Bar{N} \approx 26.6$. This disagrees with the value of $31$ quoted by the authors~\cite{trotzky_probing_2012},\footnote{31 is neither the mean, the mode (29), nor the median (which lies between 27 and 29).} but is consistent with Ref.~\cite{moussa_realizing_2017}. Here, I run simulations for all odd values of $N$ between 3 and 43. For each $N$, the initial states have the form,
\begin{equation}
    \ket{\Psi(0)} = \ket{0,0,0,\  \ldots\ 0,1,0,1,0,1,0,\ \ldots\ 0,0,0}.
\end{equation}
In other words, the central lattice sites form a charge density wave, while there are an equal number of empty sites to the left and right. I define the central, initially occupied, lattice site to have an index $j = 0$, meaning the leftmost and rightmost sites have indices $j = \pm 60$. The odd-numbered sites are thus initially unoccupied. As the particle number $N$ is conserved, I use the standard approach of employing U(1) symmetric tensors~\cite{singh_tensor_2011}. I also repartition the MPS every ten time-steps to improve load balancing.

\begin{figure}[!tb]
    \centering
    \includegraphics[width=8.5cm]{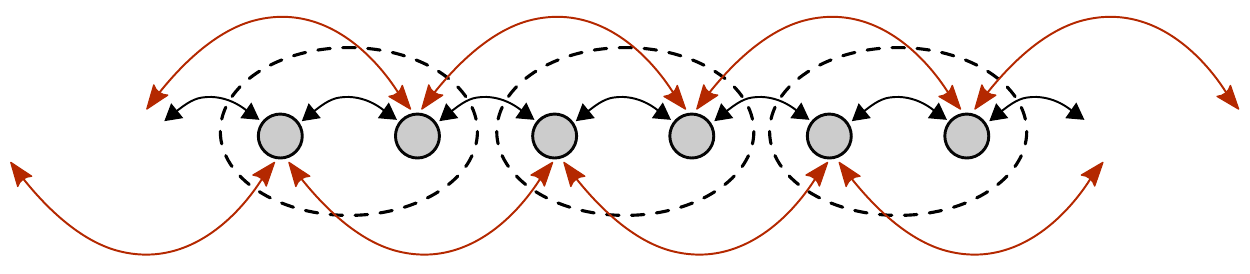}
    \colorcaption{\label{fig:grouping-sites}Schematic showing pairs of neighboring lattice sites (gray circles) grouped into ``super sites'' (dashed ellipses). This grouping converts next-nearest-neighbor hopping (long red arrows) between sites into nearest-neighbor hopping between super sites. At the same time, nearest-neighbor hopping (short black arrows) between grouped sites becomes a single super site process.}
\end{figure}

Although NNN terms can be handled in MPS simulations using SWAP gates~\cite{vidal_efficient_2003, stoudenmire_minimally_2010}, here I use the simpler, naive method of grouping every two sites into one ``super site''. As well as being simpler to implement, this approach requires fewer singular value decompositions (SVDs), so should have a smaller truncation error. On the other hand, it is likely to be less efficient as the grouping gives rise to a larger local dimension. I find a bosonic mode truncation of $n_{\text{max}} = 4$ to be sufficient, meaning I have an effective local dimension of $d_{\text{eff}} = d^2 = 25$. The grouping converts the system from a next-nearest-neighbor model of size $L$ to a nearest-neighbor model of size $L_{\text{eff}} = L/2$ (see Fig.~\ref{fig:grouping-sites}). As $L$ is odd, I add an extra empty site to the end of the chain. This has no effect on the results as the harmonic trap ensures there is a negligible probability amplitude for particles to reach the boundary sites. The resultant MPS thus consists of $L_{\text{eff}} = 61$ sites. A benefit of this approach is that every other two-site observable becomes an effective single-site observable, meaning these can be calculated in parallel as easily as the genuine single-site observables.

Following Ref.~\cite{urbanek_parallel_2016}, I use a time-step of $\delta t = 0.05 \pi/2$, and run all simulations for 100 time-steps.\footnote{M.\@ Urbanek, personal communication.} I calculate expectation values in parallel after each time-step. This comes at the cost of some additional error, as the expectation value calculations assume orthonormality. However, this error is reduced by ensuring that the kept singular values are renormalized after each truncated SVD. To converge my calculations, I run each simulation four times, using a maximum bond dimension $\chi$ of 40, 200, 1000, and 5000. I set a truncation error tolerance of ${w_\text{max} = 10^{-12}}$, and a relative truncation tolerance of ${\varepsilon = 10^{-12}}$. Convergence results are provided in the Appendix.

\subsection{Observables}\label{sec:trotzky-observables}

A subtlety of the quantum simulation experiment is that it gives access to ``quasi-local''---rather than local---observables. These are only equal in a translationally invariant setting~\cite{flesch_probing_2008}, which is not the case here. The authors first consider the quasi-local densities of the odd and even lattice sites, defined in Refs.~\cite{cramer_exploring_2008, flesch_probing_2008} as
\begin{equation}\label{eqn:quasidensities}
    n_{\text{odd}}(t) \equiv \sum_{j \, \mathrm{odd}}{n}_j(t) / N, \quad
    n_{\text{even}}(t) \equiv \sum_{j \, \mathrm{even}}{n}_j(t) / N,
\end{equation}
where $N \equiv \sum_j{n_j(t)}$ is the conserved total particle number. In addition, they track three experimental quantities related to the quasi-local nearest-neighbor correlation function
\begin{equation}\label{eqn:quasicorrelation}
    C_{\text{even}}(t) \equiv \sum_{j \, \mathrm{even}}C_{j}(t) / N,
\end{equation}
where
\begin{equation}\label{eqn:localcorrelation}
    C_{j}(t) \equiv \braket{\Psi(t)| \hat{b}_{j}^\dagger \hat{b}_{j+1} |\Psi(t)}.
\end{equation}
These are the amplitude $A(t)$ and phase $\phi(t)$~\cite[Supp.~Eqs.~1--2]{trotzky_probing_2012} of the quasi-local two-site ``tunnel oscillation''~\cite{trotzky_probing_2012},
\begin{eqnarray}\label{eqn:quasicurrent}
    A(t) e^{\mathrm{i}\phi(t)} &\equiv& \frac{1}{N} \sum_{j \, \mathrm{even}}{A}_j(t) e^{\mathrm{i}\phi_j(t)}\\* \nonumber
    &=& n_{\text{even}}(t) - n_{\text{odd}}(t) - 2 \mathrm{i} \Im\big(C_{\text{even}}(t)\big),
\end{eqnarray}
and the quasi-local ``visibility''~\cite[Supp.~Eq.~5]{trotzky_probing_2012},
\begin{eqnarray}\label{eqn:quasivisibility}
    \nu(t) &\equiv& \frac{1}{N} \sum_{j} \left(C_j(t) + C_j^*(t)\right)
    = 4 \Re\big(C_{\text{even}}(t)\big).
\end{eqnarray}
During my simulations, I compute local observables, as these provide more information than the quasi-local ones. Averaging over these gives the corresponding quasi-local quantities. At each time-step, I calculate the on-site densities
\begin{equation}\label{eqn:localdensities}
    {n}_j(t) \equiv \braket{\Psi(t)| \hat{n}_j |\Psi(t)}
\end{equation}
for every site $j$, and the real and imaginary parts of the nearest-neighbor correlation functions $C_{j}(t)$ for every other site. The physical interpretation of these quantities is as follows: $\Re\big(C_{j}(t)\big)$ is proportional to the expectation value of the energy associated with hopping between sites $j$ and $j+1$, while $\Im\big(C_{j}(t)\big)$ is proportional to the expectation value of the mass current flowing from site $j$ into site $j+1$.

\section{Results}

\begin{figure}[!t]
    \centering
    \includegraphics[width=8.5cm]{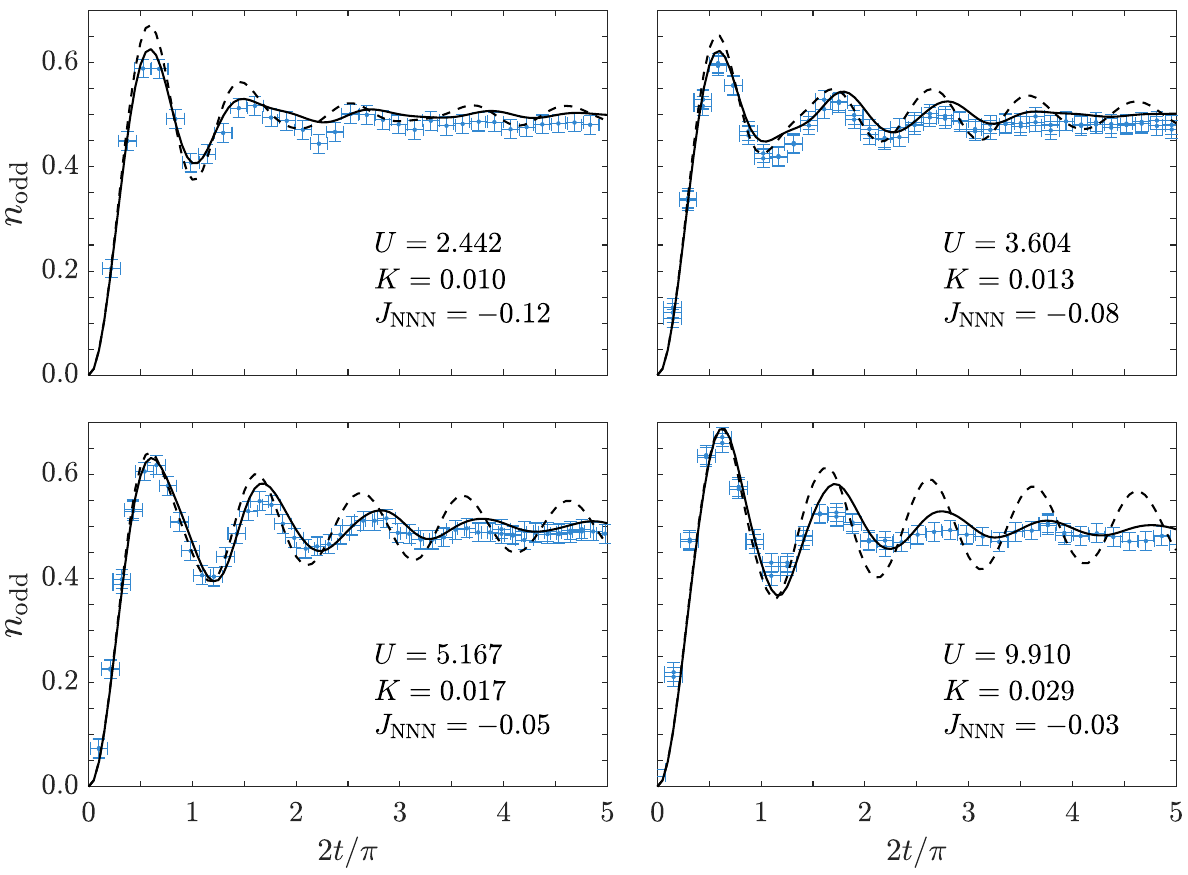}
    \colorcaption{\label{fig:trotzky-data}Relaxation of the quasi-local density $n_{\text{odd}}$. Markers are the experimental data from Ref.~\cite[Fig.~2]{trotzky_probing_2012}. Solid lines are my ensemble-averaged pTEBD results for the next-nearest-neighbor model with a maximum bond dimension of $\chi=5000$. Dashed lines are my TeNPy iTEBD2 results for the standard Bose--Hubbard model in the thermodynamic limit. The corresponding simulation parameters are given in Table~\ref{table:itebd_params}. All calculations used a time-step of $\delta t=0.05 \pi/2$.}
\end{figure}

\subsection{Comparison to experimental data}

\begin{table}[!b]
    \centering
    \begin{tabular}{cccccc} 
    \toprule
    iTEBD simulation & $U$ & $J$ & $\chi$ & $n_{\text{max}}$ & Wall time / hour \\ 
    \midrule
    a & $2.442$ & 1 & 20000 & 5 & 188.2 \\
    b & $3.604$ & 1 & 15000 & 5 & 51.6 \\
    c & $5.167$ & 1 & 10000 & 5 & 14.4 \\
    d & $9.910$ & 1 & 5000 & 5 & 2.1 \\
    \bottomrule
    \end{tabular}
    \caption{Simulation parameters and wall times for the translationally invariant iTEBD2 Bose--Hubbard model calculations. As there is no next-nearest-neighbor hopping, the local Hilbert space dimension is simply $d = n_{\text{max}} + 1 = 6$.}
    \label{table:itebd_params}
\end{table}

Fig.~\ref{fig:trotzky-data} shows my pTEBD results for the quasi-local densities, as well as the experimental results from \mbox{Ref.~\cite[Fig.~2]{trotzky_probing_2012}}.\footnote{Any error introduced by my reading of this data should be negligible compared to the experimental uncertainty.} The error bars match the size of the authors' experimental data markers, which suggest an uncertainty in $n_\text{odd}$ of $\pm 0.018$ and an uncertainty in $2t/\pi$ of $\pm 0.084$. Also shown are results for the standard Bose--Hubbard model in the thermodynamic limit (dashed curves). I calculated these using the second-order iTEBD (iTEBD2) routine in TeNPy~\cite{hauschild_efficient_2018, tenpy_developers_tensor_nodate} with the parameters shown in Table~\ref{table:itebd_params}. Notice that the pTEBD results and the experimental data both relax more rapidly than the iTEBD results. The agreement between the pTEBD and experimental results for simulation [c] is particularly striking, whereas there seems to be a small systematic error in the experimental data for simulations [a, b]. There are larger discrepancies between the pTEBD and experimental results for simulation [d]. These were already noted by Trotzky \emph{et al.}, who attribute them to ``residual inter-chain tunneling and non-adiabatic heating''~\cite{trotzky_probing_2012}.

\begin{figure}[!t]
    \centering
    \subfloat[\label{sfig:trotzky_experimental_current_A}]{
        \includegraphics[width=8.35cm]
        {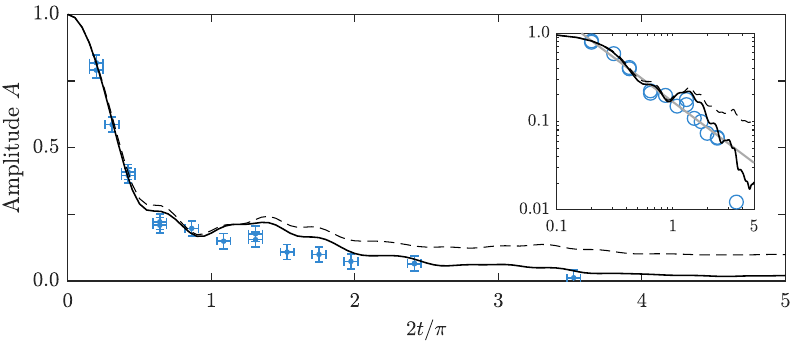}
    }\\
    \subfloat[\label{sfig:trotzky_experimental_current_phi}]{
        \includegraphics[width=8.35cm]{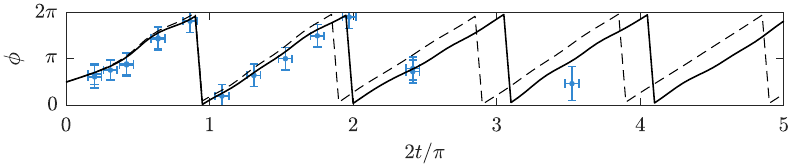}
    }\\
    \subfloat[\label{sfig:trotzky_experimental_visibility}]{
        \includegraphics[width=8.35cm]{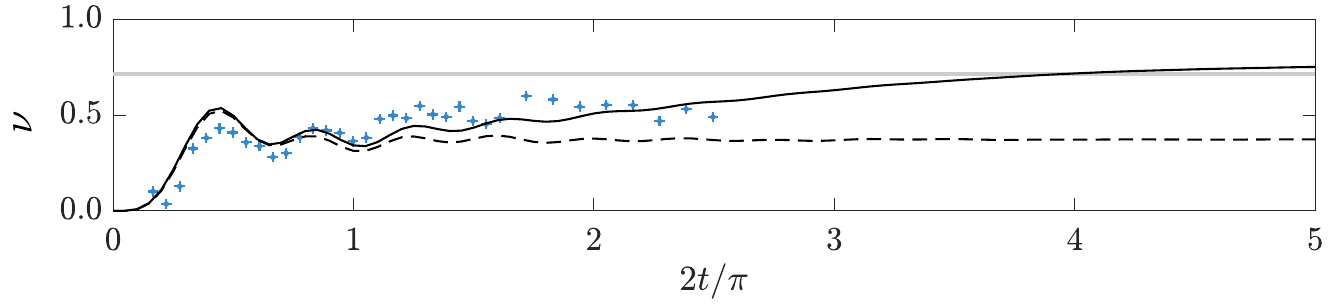}
    }
    \colorcaption{\label{fig:trotzky-5167-data}Ensemble-averaged numerics (black solid curves) versus experiment (blue markers) for simulation [c] ($U=5.167$). All experimental values are taken from Ref.~\cite[Figs.~3(c),~4(b)]{trotzky_probing_2012}, whose axis scales I attempt to match. Also shown are my iTEBD2 thermodynamic limit results for the Bose--Hubbard model (black dashed curves). \mbox{\protect\subref{sfig:trotzky_experimental_current_A} Amplitude $A$} of the quasi-local tunnel oscillation. Inset shows the same data (minus error bars) on a log-log scale. Gray solid line shows a linear fit (on log-log scale) to the experimental data, excluding the final point. \protect\subref{sfig:trotzky_experimental_current_phi} Phase $\phi$ of the quasi-local tunnel oscillation. \protect\subref{sfig:trotzky_experimental_visibility} Relaxation of the quasi-local visibility $\nu$. The gray solid line is the experimental equilibrium value of $\nu$.}
\end{figure}

\begin{figure*}
    \centering
    \includegraphics[width=11cm]{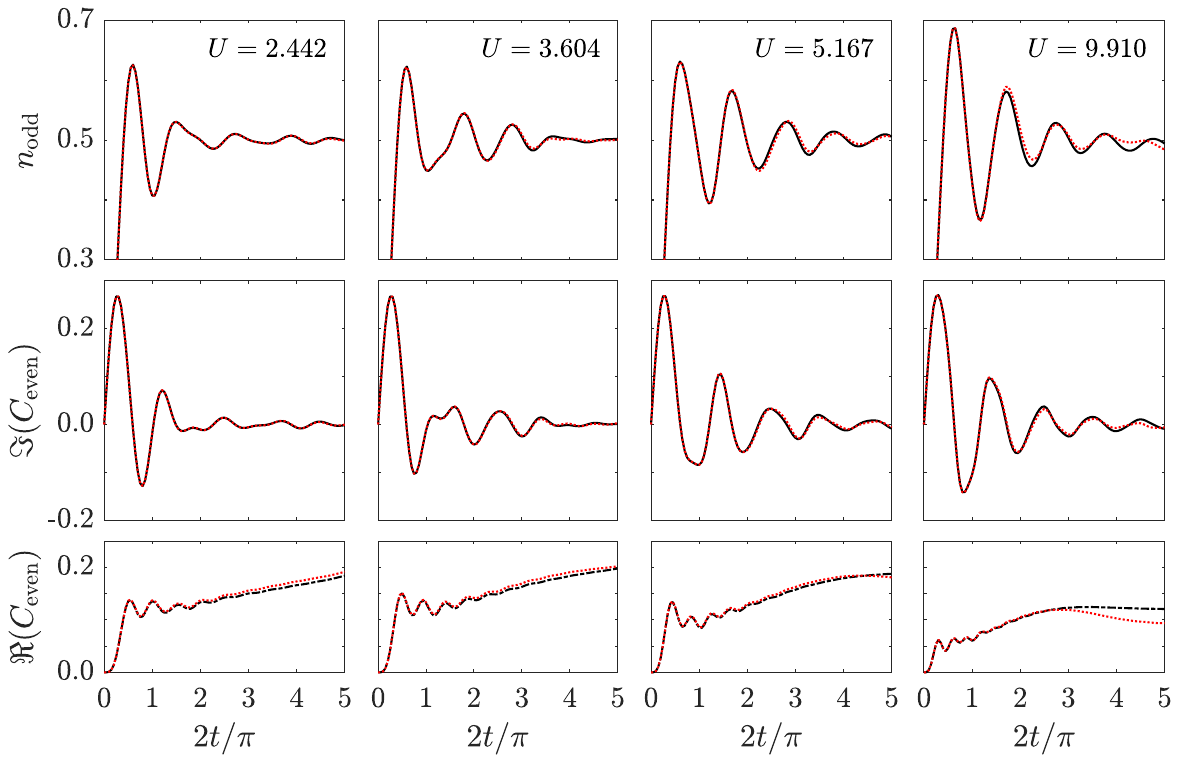}
    \colorcaption{\label{fig:ensemble-vs-27}Ensemble-averaged relaxation results (solid black curves) compared to single system results (red dotted curves) for $N=27$.}
\end{figure*}

In \subfirs{sfig:trotzky_experimental_current_A}~and~\subref*{sfig:trotzky_experimental_current_phi}, I plot the amplitude $A$ and phase $\phi$ of the quasi-local tunnel oscillation for simulation [c] ($U=5.167$), as well as the experimental results from Ref.~\cite[Fig.~3(c)]{trotzky_probing_2012}. For these observables, there is only one data point beyond the limit of author's numerics, although there is reasonable agreement with my pTEBD results. The biggest discrepancies are those already seen in Ref.~\cite[Fig.~3(c)]{trotzky_probing_2012} for $A$ between times $1 < 2t/\pi < 2$. The dashed curves are the NN thermodynamic limit results calculated with iTEBD, which again give a less accurate description of the experiment. In order to replicate the authors' power-law fit for $A$, I perform a linear least-squares fit to the experimental data on a log-log scale (solid gray line shown in the inset of \subref*{sfig:trotzky_experimental_current_A}). However, to match the fit of Ref.~\cite[Fig.~3(c) (left inset)]{trotzky_probing_2012}, I find I must exclude the final experimental data point. The resultant line gives a good fit to the other experimental data points but does not agree as well with the numerics.

Finally, in \subfir{sfig:trotzky_experimental_visibility}, I compare my results for the quasi-local visibility to the experimental data from Ref.~\cite[Fig.~4(b)]{trotzky_probing_2012}. I find the relaxed value of this observable just slightly larger than the experimental value. This small disagreement is likely due to experimental uncertainty. I also show the visibility for the NN thermodynamic limit case calculated with iTEBD (dashed curve). Notice that this relaxes more quickly to approximately half the experimental value, as also noted by Trotzky \emph{et al.\@} (compare Ref.~\cite[Fig.~4(c)]{trotzky_probing_2012} to Ref.~\cite[Fig.~13]{flesch_probing_2008}).

\subsection{Effect of ensemble averaging}\label{sec:trotzky-ensemble-averaging}

To corroborate the claim that the fast relaxation cannot be accounted for by the ensemble averaging~\cite{trotzky_probing_2012}, I compare the averaged results to those for a single chain. The numerics in Fig.~\ref{fig:ensemble-vs-27} show that the effect of the ensemble averaging is mostly insignificant over the simulated timescale. In fact, the relaxation behavior of the ensemble-averaged observables is very close to that of the single $N=27 \approx \Bar{N}$ chain---the only exception being simulation [d] ($U=9.910$), where the real part of $C_{\text{even}}$ starts to deviate after $2t/\pi \approx 3$ and $n_{\text{odd}}$ starts to deviate after $2t/\pi \approx 4$. However, I find the ensemble-averaged results to deviate more significantly from those of the single $N=43$ chain considered in Ref.~\cite[Supp.~Fig.~2]{trotzky_probing_2012}.

\subsection{Effect of next-nearest-neighbor hopping}\label{sec:nnn-hopping}

\begin{figure}[!b]
    \centering
    \includegraphics[width=5.4cm]{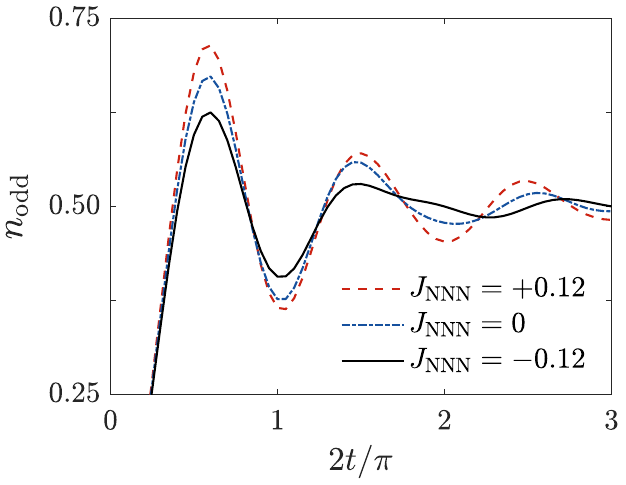}
    \colorcaption{\label{fig:Jnnn-sign-comparison}Effect of the next-nearest-neighbor hopping amplitude $J_{\mathrm{NNN}}$ on the relaxation of the quasi-local density $n_{\text{odd}}$ for the $U=2.442$, $K=0.010$ case with $N=27$ bosons. These data were calculated with $\chi=1000$.}
\end{figure}

As noted in Ref.~\cite{trotzky_probing_2012}, the effect of the next-nearest-neighbor hopping is most prominent for simulations [a, b]. In Fig.~\ref{fig:Jnnn-sign-comparison}, I plot the evolution of the quasi-local density for simulation [a] with $J_{\mathrm{NNN}}$ positive, negative, and zero. Notice that a positive value leads to slower relaxation, whereas a negative value leads to the faster relaxation observed by Trotzky \emph{et al.} I find a similar effect for the relaxation of the currents, while the effect on the visibilities is relatively small (deviations of less than $10\%$ in all simulations).

\subsection{Effect of harmonic trap}

\begin{figure}[!p]
    \centering
    \includegraphics[width=8.45cm]{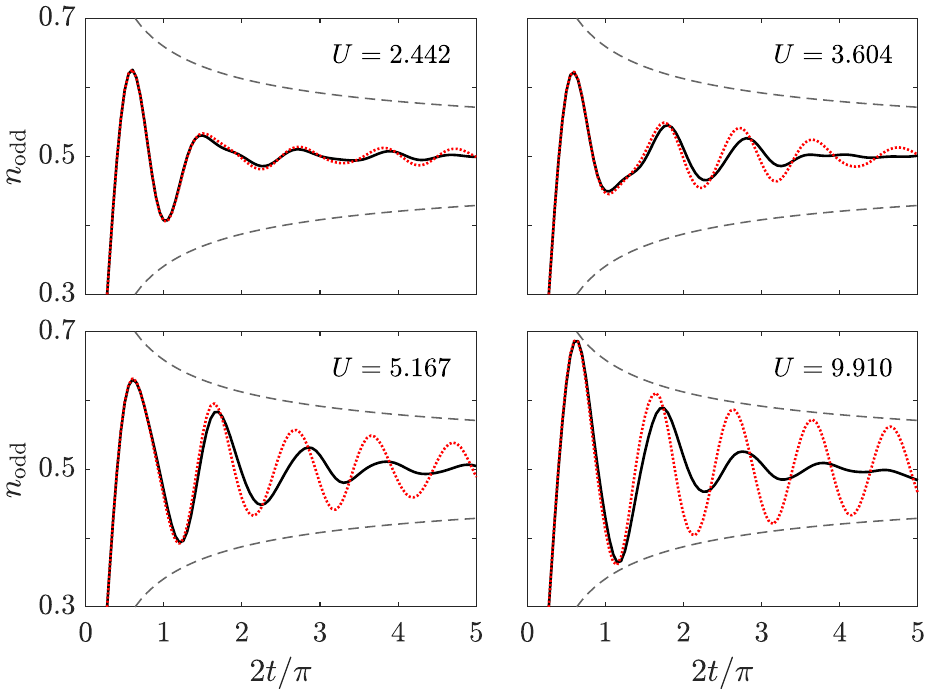}
    \colorcaption{\label{fig:trap-vs-no-trap}Evolution of quasi-local densities $n_{\text{odd}}$ for $N=27$ with trap present (black solid curves) and in its absence (red dotted curves). Gray dashed curves are the envelope for the case of the free thermodynamic limit Bose--Hubbard model.}
\end{figure}
\begin{figure}[!p]
    \centering
    \subfloat[\label{sfig:density_U-9910_N-27}]{
    \includegraphics[width=8.35cm]{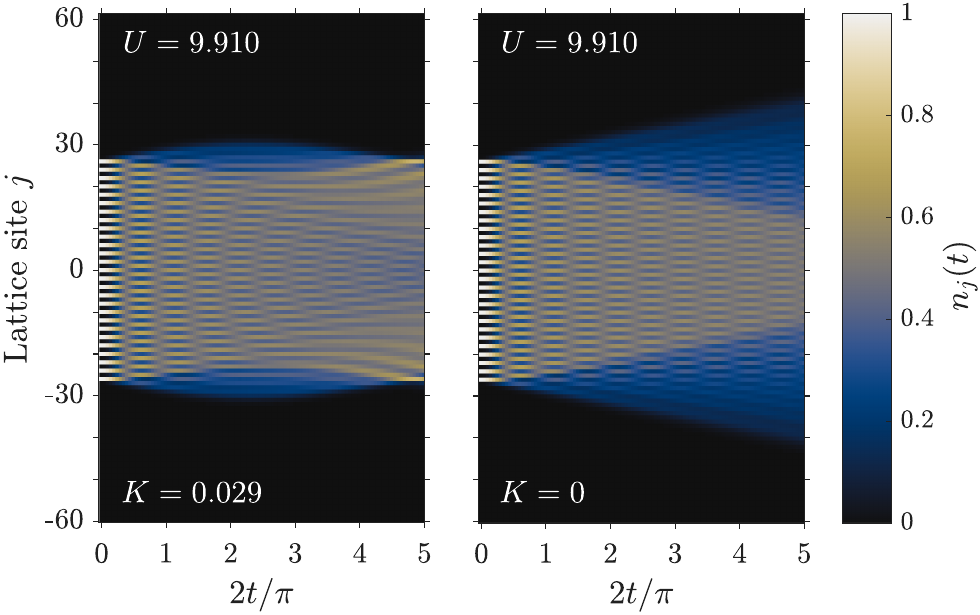}
    }\\
    \subfloat[\label{sfig:density_U-9910_N-43}]{
    \includegraphics[width=8.35cm]{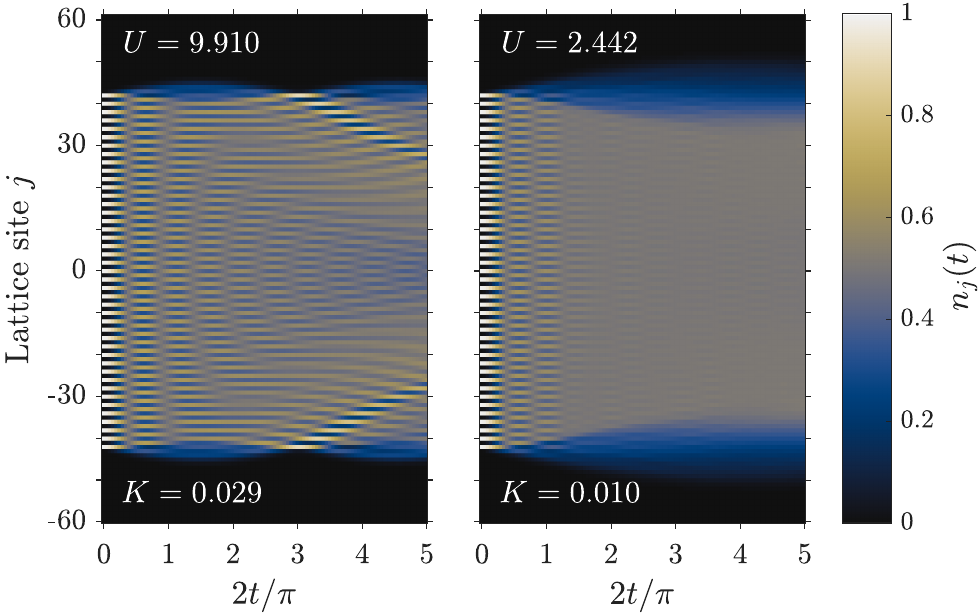}
    }
    \colorcaption{\label{fig:density-plot-trap-comparison}Relaxation of the local densities $n_j$. \protect\subref{sfig:density_U-9910_N-27} $N=27$ CDW with $U=9.910$, $J_{\text{NNN}}=-0.03$. Left panel shows bosons confined in a $K=0.029$ harmonic trap, while right panel shows their ballistic expansion for $K=0$. \protect\subref{sfig:density_U-9910_N-43} $N=43$ CDW with (left) $U=9.910$, $J_{\text{NNN}}=-0.03, K=0.029$, and (right) $U=2.442$, $J_{\text{NNN}}=-0.12, K=0.010$.}
\end{figure}

I next investigate the effect of the harmonic trapping potential. In contrast to the small effect of the ensemble averaging, the relaxation of the quasi-local observables can be greatly enhanced by the presence of the harmonic trap~\cite{secular_comment_2020}. In Fig.~\ref{fig:trap-vs-no-trap}, for example, I show the effect on the quasi-local densities. As can be seen, the fast relaxation observed in simulations [c, d] is largely down to the trapping potential. Note that this is consistent with the results of Ref.~\cite[Fig.~10]{flesch_probing_2008} for the $U=0$ (or equivalently $U \to \infty$) Bose--Hubbard model.

To understand this, consider the left panel of \subfir{sfig:density_U-9910_N-27}. Notice that the soft reflections from the trap cause the local density oscillations to go out of phase with each other. This leads to an artificial decay of the quasi-local densities. Contrast this with the right panel of \subfir{sfig:density_U-9910_N-27} where I show the evolution in the absence of a trap. As expected, the effect of the trap grows with increasing $K$ and $N$, as these effectively reduce the system size. The enhanced relaxation is most pronounced for simulation~[d] and smallest for simulation [a] (see \subfir{sfig:density_U-9910_N-43}), as these have the strongest ($K = 0.029$) and weakest ($K = 0.010$) traps, respectively.

\begin{figure}[!tb]
    \centering
    \includegraphics[width=8.4cm]{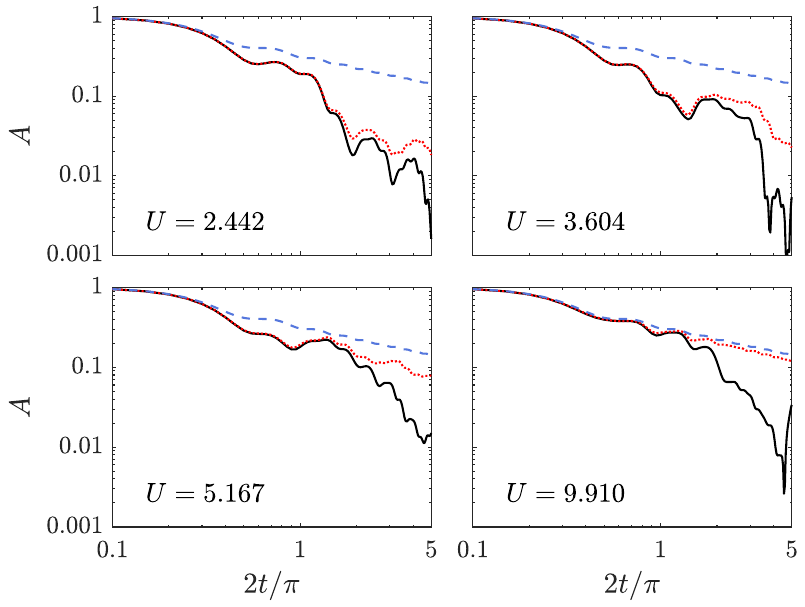}
    \colorcaption{\label{fig:A-trap-vs-no-trap}Quasi-local tunnel oscillation amplitude $A$ versus time for $N=27$ with trap (black solid curves) and without (red dotted curves). Dashed blue curves are the exact thermodynamic limit result for the free Bose--Hubbard model.}
\end{figure}

The effect of the harmonic trap can also been seen vividly in the decay of the tunnel oscillation amplitude $A$. To demonstrate this, I plot the evolution of $A$ on a log-log scale in Fig.~\ref{fig:A-trap-vs-no-trap}. A clear difference is visible between the behavior in the presence of the trap (black solid curves) and its absence (red dotted curves) for all simulations after about $2t/\pi = 2$. However, the effect is again largest for simulation [d]: with no trap, the evolution of $A$ is very close to that of the free NN Bose--Hubbard model (blue dashed curve), with its asymptotic power-law decay. For smaller values of $U$, the behavior is not well described by the free model. In fact, the decay seems to follow neither a simple power law nor exponential over this timescale (the asymptotic behavior remains an open question).

\section{Parallel performance}

\begin{figure}[!t]
    \centering
    \includegraphics[width=8.5cm]{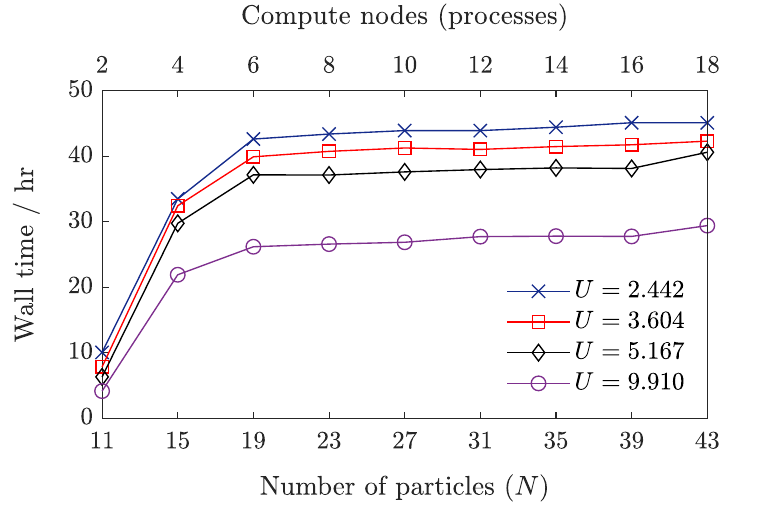}
    \colorcaption{\label{fig:trotzky_weak_scaling}Weak scaling plot showing total wall time in hours versus number of particles $N$ and number of MPI processes $p$, where $p = (N-7)/2$. All calculations used a maximum bond dimension of $\chi = 5000$. Note that the wall times are inclusive of all expectation value calculations, repartitioning, file output, and other overheads.}
\end{figure}

Due to the large dimensionality of the problem, a strong scaling analysis is impractical as the calculations would likely take a few weeks to run on a single process. However, a weak scaling analysis is possible, with the problem size taken as the number of particles $N$. For a fixed maximum bond dimension, the computation time should grow approximately linearly with $N$. For $N \geq 11$, I run the simulations on an even number of processes $p$, with each process assigned to a separate compute node. I choose $p$ as a function of $N$ given by
\begin{equation}
    p = \begin{cases}
        (N-7)/2,\qquad \text{if } (N+1)/2 \text{ is even}, \\
        (N-9)/2,\qquad \text{if } (N+1)/2 \text{ is odd}.
    \end{cases} 
\end{equation}
For $N<11$, parallelization was unnecessary; these calculations were carried out sequentially on a single compute node. In Fig.~\ref{fig:trotzky_weak_scaling}, I present weak scaling results for $p = (N-7)/2$. All calculations (including those not shown) took less than 50 hours with a maximum bond dimension of $\chi = 5000$. Notice that for $N < 15$, the calculations are much faster. This can be understood from the fact that $\chi$ is always saturated at the same simulation time for $N \geq 15$, whereas it is saturated at a later time (if at all) for $N \leq 11$. An appropriate comparison of the wall times is thus only possible for $N$ between 15 and 43. In this regime, I find fairly good weak scaling, with the wall times constant to within a few percent for $19 \leq N \leq 39$ (corresponding to $6 \leq p \leq 16$).

\section{Discussion}

In this work I have shown how parallel TEBD makes feasible the verification of an analog 1D bosonic quantum simulator in the classically accessible regime. I also used infinite TEBD to simulate the corresponding translationally invariant system, noting that it is unable to capture the effects of a harmonic trapping potential. While I could not reach the experimental timescale of $2t/\pi=20$ (equivalent to 20 ms), due to the rapid growth of entanglement, I found that it was possible to accurately compute quasi-local densities and correlation functions up to a time of $2t/\pi = 5$ (i.e.\@ 5 ms), which is the point at which they appear ``fully relaxed''~\cite{trotzky_probing_2012}. Extending the numerics of Ref.~\cite{trotzky_probing_2012} in this way allowed me to check the authors' experimental results~\cite[Figs.~2, 3(c), 4(b)]{trotzky_probing_2012}. I went on to study the effects of the harmonic trap, next-nearest-neighbor hopping, and ensemble averaging, making two observations: firstly, that the trapping potential plays a significant role in the observed relaxation for some sets of experimental parameters, and secondly that there is a crossover from the asymptotic power-law decay seen in the non-interacting model to a more complex relaxation dynamics.

In my simulations, I accounted for the next-nearest-neighbor hopping by grouping pairs of sites of dimension $d$ into super sites of dimension $d^2$. This allowed me to treat the system as a nearest-neighbor model, amenable to simulation by standard pTEBD. It also allowed me to calculate all expectation values in parallel. However, it may be more efficient to retain the $d$-dimensional sites and use SWAP gates to handle the longer-ranged terms---although this would involve truncating twice as many bonds, so may incur a larger error. For the next-nearest-neighbor case discussed here, this approach should be possible to parallelize, albeit with additional operations required at partition boundaries, which would affect the parallel scaling. Nearest-neighbor expectation values could also be carried out in parallel, but again would involve an additional delay at partition boundaries.

Using algorithmic improvements to reduce the overhead of the SVD~\cite{tamascelli_improved_2015, unfried_fast_2023} should allow a larger bond dimension (e.g.\@ $\chi \sim 25000$) to be reached, and hence a slightly larger timescale. Employing GPUs or TPUs, a bond dimension of $\chi \sim 125000$ may even be feasible~\cite{ganahl_density_2023}. However, any TEBD-type simulation will soon hit an exponential wall. As an alternative approach, one could instead try switching to the single-site TDVP algorithm (1TDVP). As single-site DMRG has already been parallelized~\cite{stoudenmire_real-space_2013, depenbrock_tensor_2013}, it should be straightforward to generalize the parallelization scheme~\cite{secular_parallel_2020} to 1TDVP. Although this would likely have a smaller time-step error, and would incur no further truncation error, it would introduce an \emph{a priori} unknown projection error~\cite{goto_performance_2019}. A numerical investigation of this approach could be an interesting avenue for future work.

\begin{acknowledgments}
The author thanks Stephen R.\@ Clark, Matthew W.\@ Cook, Jonathan Coulthard, Lennart Dabelow, \nobreak{Sergey} \nobreak{Dolgov}, Ben Fowler, Johannes Hauschild, \nobreak{Junjie} \nobreak{Liu}, and Miroslav Urbanek for helpful discussions, and gratefully acknowledges the University of Bath’s \nobreak{Research} Computing Group for their support in this work~\cite{noauthor_university_2018}. This research made use of the \nobreak{Balena} High \nobreak{Performance} \nobreak{Computing} (HPC) Service~\cite{university_of_bath_balena_nodate} with funding provided by \nobreak{ClusterVision} and the University of Bath. The author's pTEBD implementation was based on TNT Library code written by Sarah Al-Assam and Chris Goodyer~\cite{al-assam_tensor_2017, goodyer_tnt_2013}.
\end{acknowledgments}


\appendix*

\section{numerical convergence}

In this appendix, I empirically justify my parallel TEBD (pTEBD) results by considering how they depend on time-step $\delta t$ and maximum bond dimension $\chi$.

\subsection{Time-step}

\begin{figure}[!b]
    \centering
    \includegraphics[width=6.4cm]{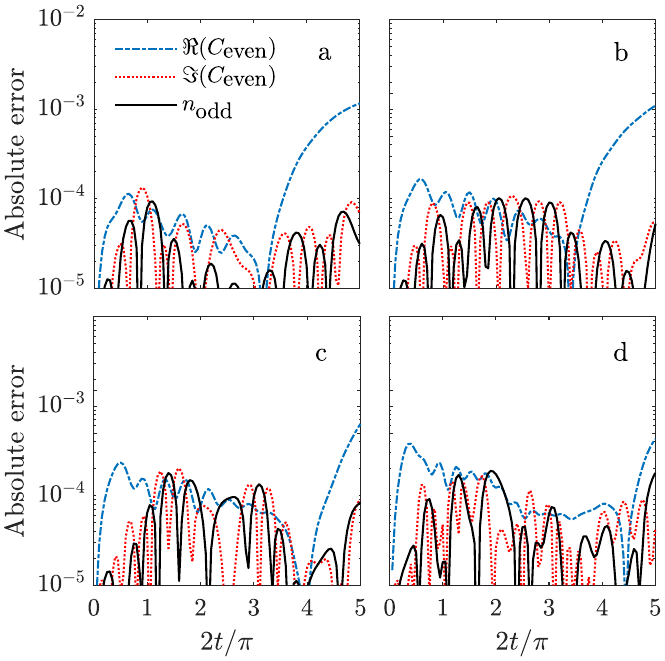}
    \colorcaption{\label{fig:trotzky-dt-errors}Comparison of second- and fourth-order pTEBD results for $N=27$. Fourth-order results were calculated sequentially using TeNPy.}
\end{figure}

To check that the chosen value of $\delta t = 0.05 \pi/2$ is small enough, I compare my second-order pTEBD results for $N=27$, $\chi = 1000$ to fourth-order pTEBD calculations with the same bond dimension and time-step. I carry out these latter simulations using the open source TeNPy~\cite{hauschild_efficient_2018, tenpy_developers_tensor_nodate} software, allowing me to demonstrate consistency between independent codebases. This is a particularly important test given that I am unable to reproduce the $t$-DMRG results of Ref.~\cite{trotzky_probing_2012}---neither their ensemble-averaged results~\cite{trotzky_probing_2012} nor their single $N=43$ chain results~\cite[Supp.~Fig.~2]{trotzky_probing_2012}---despite checking both the next-nearest-neighbor (NNN) and the nearest-neighbor (NN) cases.\footnote{Note, however, that I am able to reproduce the NN results of Refs.~\cite{flesch_probing_2008, urbanek_parallel_2016}.}

\begin{figure}[!b]
    \centering
    \includegraphics[width=8.6cm,keepaspectratio]{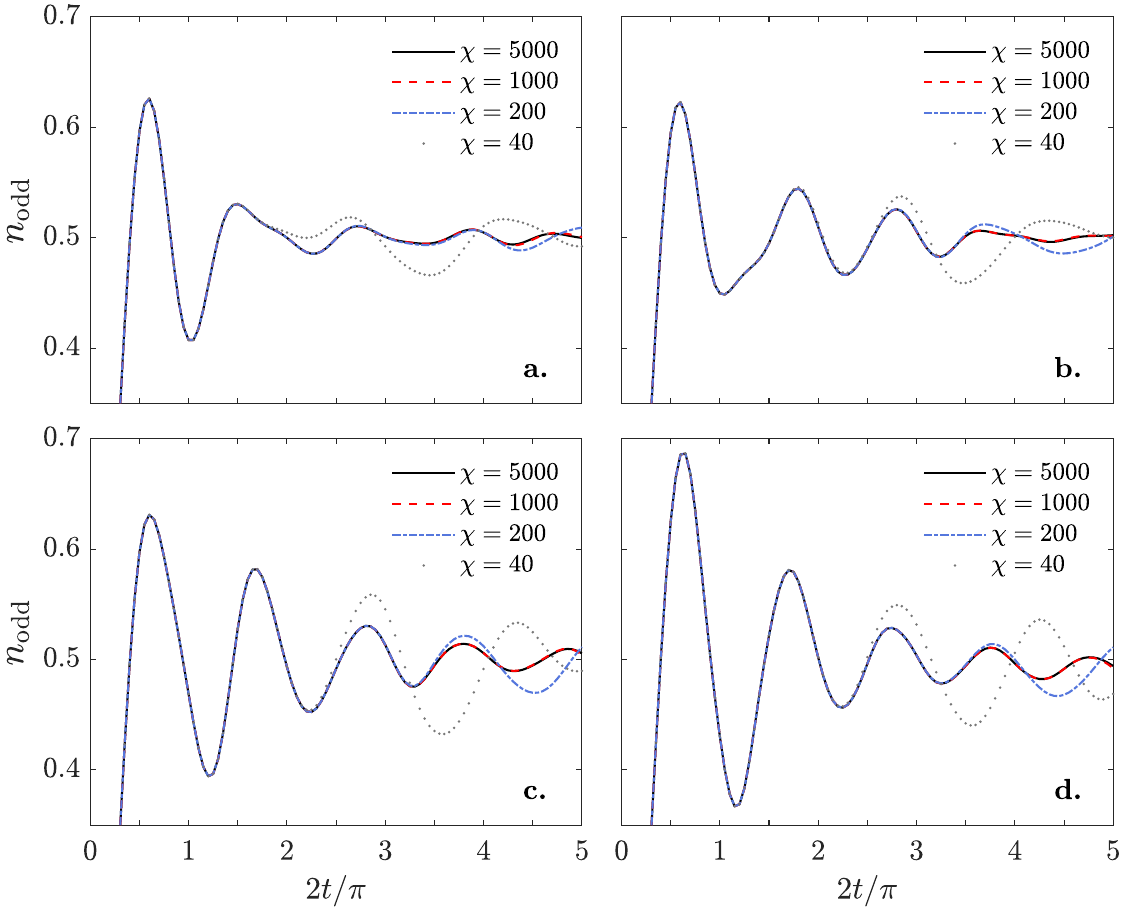}
    \colorcaption{\label{fig:n_odd-convergence}Convergence of ensemble-averaged quasi-local densities $n_{\text{odd}}$ with maximum bond dimension $\chi$.}
\end{figure}

\begin{figure}[!b]
    \captionsetup[subfloat]{position=top,labelformat=empty}
    \centering
    \subfloat{
    \includegraphics[width=8.4cm]{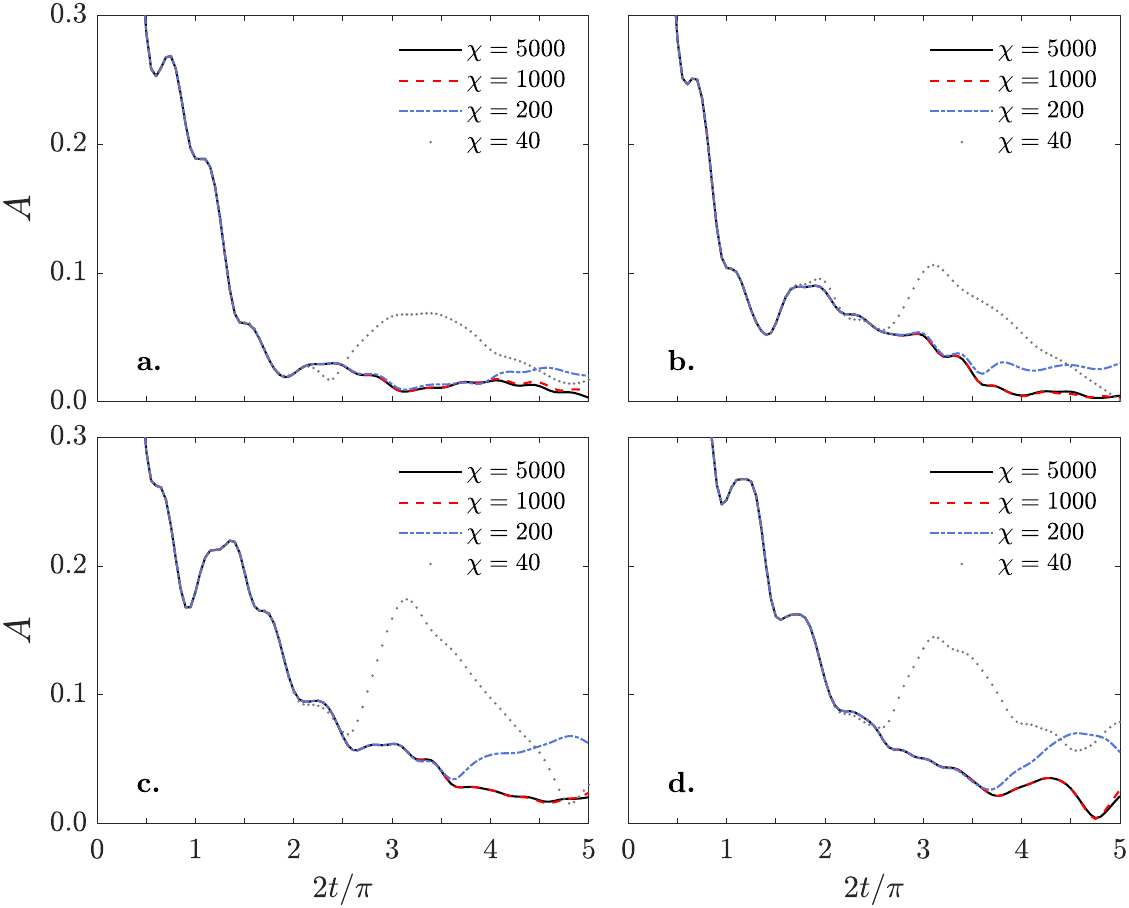}
    }\\
    \subfloat{
    \includegraphics[width=8.4cm]{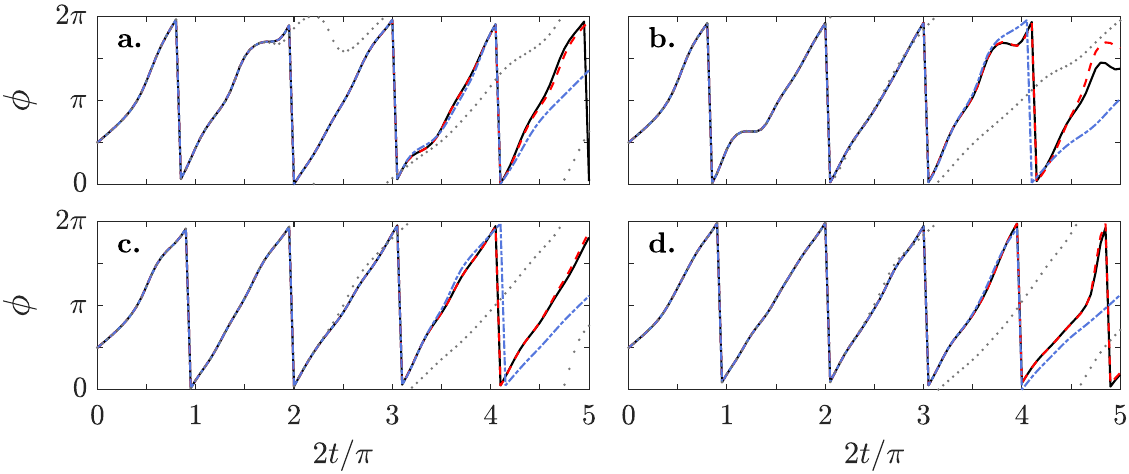}
    }
    \colorcaption{\label{fig:current-convergence}Convergence of ensemble-averaged quasi-local tunnel oscillations (amplitude $A$ and phase $\phi$) with maximum bond dimension $\chi$.}
\end{figure}

As shown in Fig.~\ref{fig:trotzky-dt-errors}, absolute differences in $n_{\text{odd}}$ and $\Im(C_{\text{even}})$ are of the order of $10^{-4}$. Differences in $\Re(C_{\text{even}})$ are of the same order up until some time $2t/\pi > 3$, when they grow monotonically by up to an order of magnitude. This is characteristic of ``runaway'' truncation error~\cite{gobert_real-time_2005} and is likely due to the small bond dimension used for these calculations.

\subsection{Bond dimension}

I next consider convergence with respect to $\chi$. In Fig.~\ref{fig:n_odd-convergence}, I show the convergence of $n_{\text{odd}}$. The fast convergence of this observable means that even a bond dimension of $\chi=1000$ gives greater precision than the quantum simulator. It is also clear that the timescale reached numerically by Trotzky \emph{et al.\@}~\cite{trotzky_probing_2012} can be reproduced using a maximum bond dimension of $\chi = 200$. In fact, a maximum bond dimension of just $\chi = 40$ is sufficient to reach a time of $2t/\pi = 2$.

\begin{figure}[!tbp]
    \centering
    \includegraphics[width=8.5cm]{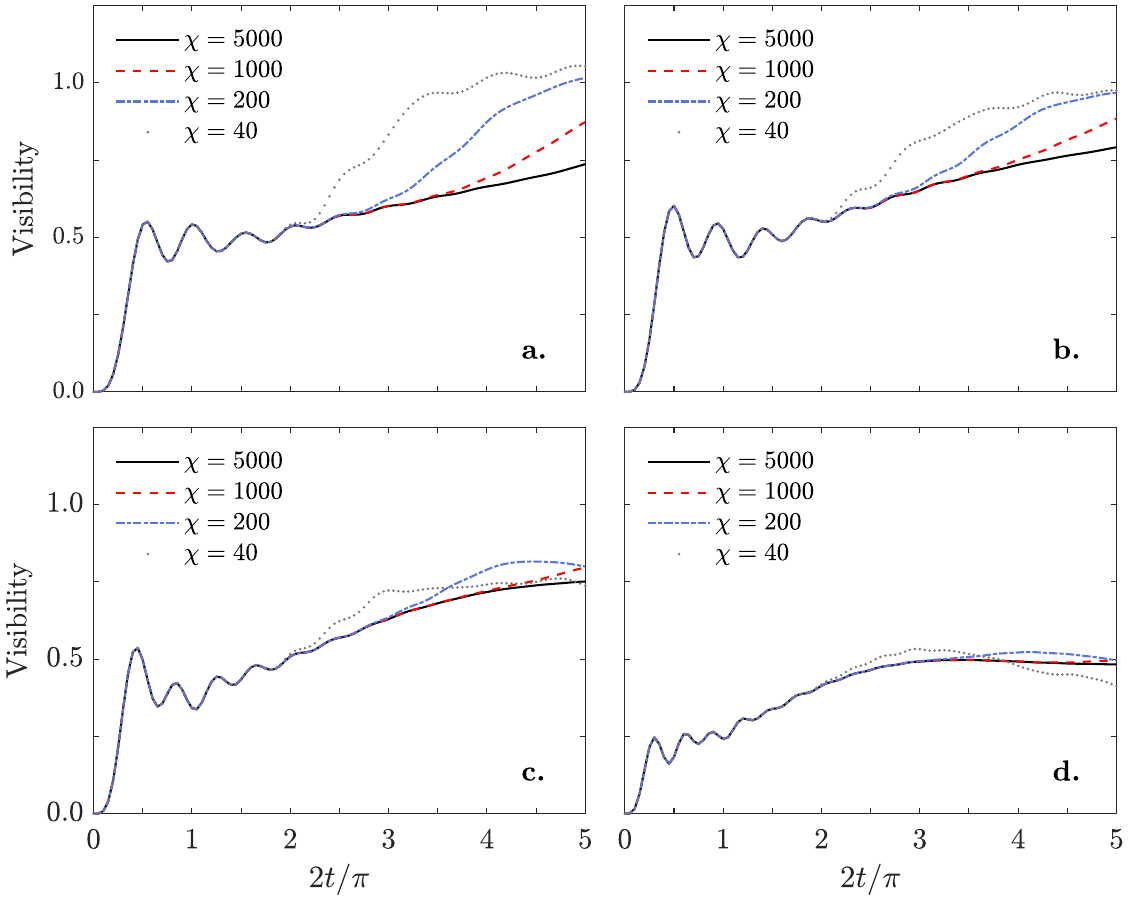}
    \colorcaption{\label{fig:vis-convergence}Convergence of the ensemble-averaged quasi-local visibilities $\nu = 4 \Re(C_{\text{even})}$ with maximum bond dimension $\chi$.}
\end{figure}

The convergence of $A$ and $\phi$ is shown in Fig.~\ref{fig:current-convergence}. These observables also converge fairly quickly, though deviations between the $\chi=1000$ and $\chi=5000$ results are visible in simulations [a, b] for $2t/\pi \gtrsim 4$.

\begin{figure}[!b]
    \centering
    \includegraphics[width=8.5cm]{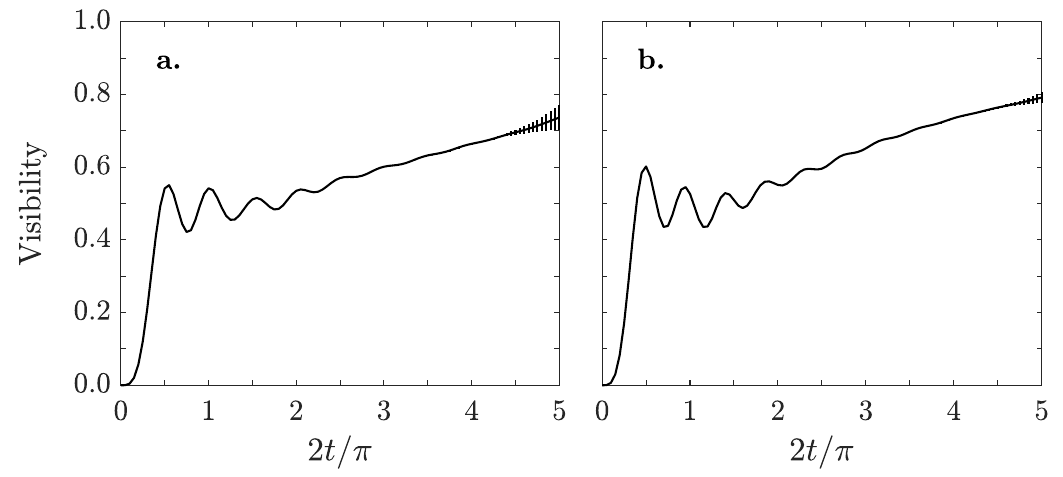}
    \caption{Ensemble-averaged visibility for simulations [a, b] with $\chi=5000$. The error bars show the estimated error of $4 \epsilon_{\text{local}}$, where $\epsilon_{\text{local}}$ is the scaled total discarded weight from the $N=27$ calculations (see Fig.~\ref{fig:trotzky-discarded-weights}). As the truncation causes the visibilities to overshoot for smaller values of $\chi$ (see Fig.~\ref{fig:vis-convergence}), I would expect the correct values to lie inside the lower error bars.}
    \label{fig:trotzky-visibilities-errorbars}
\end{figure}

Finally, in Fig.~\ref{fig:vis-convergence}, I show the visibility $\nu$, whose convergence is slower than that of the corresponding currents and densities. Indeed, for simulations [a, b], it is not clear whether the calculations have converged for $2t/\pi \gtrsim 4.25$. I therefore calculate error bars as shown in Fig.~\ref{fig:trotzky-visibilities-errorbars} for these two cases. To understand how the errors were estimated, first note that the total discarded weight is a global measure of the error in the quantum state. It typically provides far too loose a bound on the error in local (or quasi-local) quantities. Empirically, I find a better bound is given by the \emph{scaled} total discarded weight $\epsilon_{\text{local}} = 2 w_{\mathrm{total}} /N$.

\begin{figure}[!b]
    \centering
    \includegraphics[width=8.5cm]{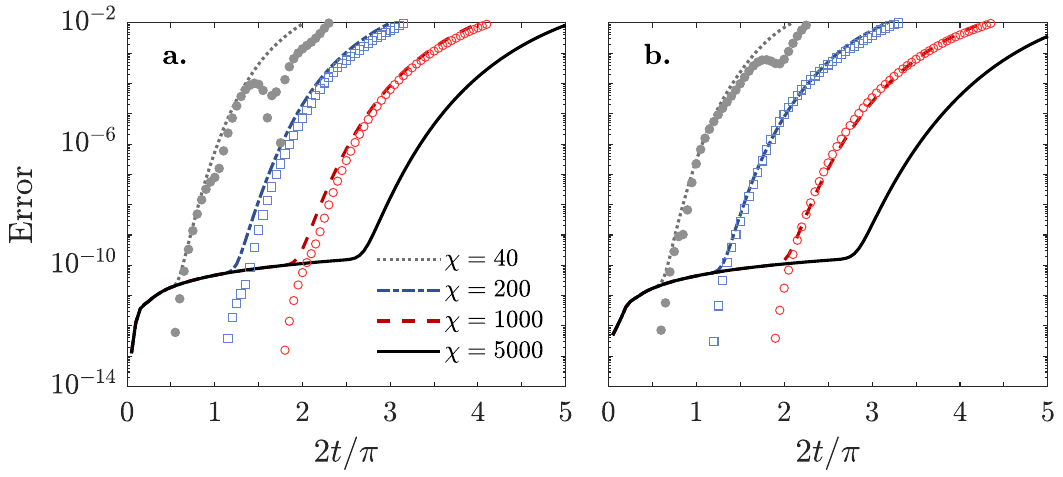}
    \colorcaption{\label{fig:trotzky-discarded-weights}Error measures for simulations {[a, b]} versus time. The curves show the scaled total discarded weights $\epsilon_{\text{local}} = 2 w_{\mathrm{total}} /N$ for the $N=27$ case. The markers show the absolute differences in the ensemble-averaged values of $\Re(C_{\text{even})}$ as compared to the $\chi=5000$ reference case.}
\end{figure}

To numerically estimate the error in an observable simulated using $\chi=40, 200, 1000$, I use the $\chi=5000$ calculation as a reference and calculate the absolute difference. Fig.~\ref{fig:trotzky-discarded-weights} shows the evolution of this absolute error in the ensemble-averaged $\Re(C_{\text{even}})$ for simulations [a, b]. In the same figure, I plot $\epsilon_{\text{local}}$ for the simulations with $N=27$. These quantities closely match, suggesting that the scaled total discarded weight can be used as a proxy for the error in this observable. I thus use $\epsilon_{\text{local}}$ for the $N=27$, $\chi=5000$ simulation to estimate the error bars shown in Fig.~\ref{fig:trotzky-visibilities-errorbars}.

\bibliography{parallel_tebd_case_study.bib}

\begin{thebibliography}{38}%
\makeatletter
\providecommand \@ifxundefined [1]{%
 \@ifx{#1\undefined}
}%
\providecommand \@ifnum [1]{%
 \ifnum #1\expandafter \@firstoftwo
 \else \expandafter \@secondoftwo
 \fi
}%
\providecommand \@ifx [1]{%
 \ifx #1\expandafter \@firstoftwo
 \else \expandafter \@secondoftwo
 \fi
}%
\providecommand \natexlab [1]{#1}%
\providecommand \enquote  [1]{``#1''}%
\providecommand \bibnamefont  [1]{#1}%
\providecommand \bibfnamefont [1]{#1}%
\providecommand \citenamefont [1]{#1}%
\providecommand \href@noop [0]{\@secondoftwo}%
\providecommand \href [0]{\begingroup \@sanitize@url \@href}%
\providecommand \@href[1]{\@@startlink{#1}\@@href}%
\providecommand \@@href[1]{\endgroup#1\@@endlink}%
\providecommand \@sanitize@url [0]{\catcode `\\12\catcode `\$12\catcode `\&12\catcode `\#12\catcode `\^12\catcode `\_12\catcode `\%12\relax}%
\providecommand \@@startlink[1]{}%
\providecommand \@@endlink[0]{}%
\providecommand \url  [0]{\begingroup\@sanitize@url \@url }%
\providecommand \@url [1]{\endgroup\@href {#1}{\urlprefix }}%
\providecommand \urlprefix  [0]{URL }%
\providecommand \Eprint [0]{\href }%
\providecommand \doibase [0]{https://doi.org/}%
\providecommand \selectlanguage [0]{\@gobble}%
\providecommand \bibinfo  [0]{\@secondoftwo}%
\providecommand \bibfield  [0]{\@secondoftwo}%
\providecommand \translation [1]{[#1]}%
\providecommand \BibitemOpen [0]{}%
\providecommand \bibitemStop [0]{}%
\providecommand \bibitemNoStop [0]{.\EOS\space}%
\providecommand \EOS [0]{\spacefactor3000\relax}%
\providecommand \BibitemShut  [1]{\csname bibitem#1\endcsname}%
\let\auto@bib@innerbib\@empty
\bibitem [{\citenamefont {Daley}\ \emph {et~al.}(2022)\citenamefont {Daley}, \citenamefont {Bloch}, \citenamefont {Kokail}, \citenamefont {Flannigan}, \citenamefont {Pearson}, \citenamefont {Troyer},\ and\ \citenamefont {Zoller}}]{daley_practical_2022}%
  \BibitemOpen
  \bibfield  {author} {\bibinfo {author} {\bibfnamefont {A.~J.}\ \bibnamefont {Daley}}, \bibinfo {author} {\bibfnamefont {I.}~\bibnamefont {Bloch}}, \bibinfo {author} {\bibfnamefont {C.}~\bibnamefont {Kokail}}, \bibinfo {author} {\bibfnamefont {S.}~\bibnamefont {Flannigan}}, \bibinfo {author} {\bibfnamefont {N.}~\bibnamefont {Pearson}}, \bibinfo {author} {\bibfnamefont {M.}~\bibnamefont {Troyer}},\ and\ \bibinfo {author} {\bibfnamefont {P.}~\bibnamefont {Zoller}},\ }\bibfield  {title} {{\selectlanguage {en}\bibinfo {title} {Practical quantum advantage in quantum simulation}},\ }\href {https://doi.org/10.1038/s41586-022-04940-6} {\bibfield  {journal} {\bibinfo  {journal} {Nature}\ }\textbf {\bibinfo {volume} {607}},\ \bibinfo {pages} {667} (\bibinfo {year} {2022})}\BibitemShut {NoStop}%
\bibitem [{\citenamefont {Hoefler}\ \emph {et~al.}(2023)\citenamefont {Hoefler}, \citenamefont {Häner},\ and\ \citenamefont {Troyer}}]{hoefler_disentangling_2023}%
  \BibitemOpen
  \bibfield  {author} {\bibinfo {author} {\bibfnamefont {T.}~\bibnamefont {Hoefler}}, \bibinfo {author} {\bibfnamefont {T.}~\bibnamefont {Häner}},\ and\ \bibinfo {author} {\bibfnamefont {M.}~\bibnamefont {Troyer}},\ }\bibfield  {title} {\bibinfo {title} {Disentangling {Hype} from {Practicality}: {On} {Realistically} {Achieving} {Quantum} {Advantage}},\ }\href {https://doi.org/10.1145/3571725} {\bibfield  {journal} {\bibinfo  {journal} {Communications of the ACM}\ }\textbf {\bibinfo {volume} {66}},\ \bibinfo {pages} {82} (\bibinfo {year} {2023})}\BibitemShut {NoStop}%
\bibitem [{\citenamefont {Shaffer}\ \emph {et~al.}(2021)\citenamefont {Shaffer}, \citenamefont {Megidish}, \citenamefont {Broz}, \citenamefont {Chen},\ and\ \citenamefont {Häffner}}]{shaffer_practical_2021}%
  \BibitemOpen
  \bibfield  {author} {\bibinfo {author} {\bibfnamefont {R.}~\bibnamefont {Shaffer}}, \bibinfo {author} {\bibfnamefont {E.}~\bibnamefont {Megidish}}, \bibinfo {author} {\bibfnamefont {J.}~\bibnamefont {Broz}}, \bibinfo {author} {\bibfnamefont {W.-T.}\ \bibnamefont {Chen}},\ and\ \bibinfo {author} {\bibfnamefont {H.}~\bibnamefont {Häffner}},\ }\bibfield  {title} {{\selectlanguage {en}\bibinfo {title} {Practical verification protocols for analog quantum simulators}},\ }\href {https://doi.org/10.1038/s41534-021-00380-8} {\bibfield  {journal} {\bibinfo  {journal} {npj Quantum Information}\ }\textbf {\bibinfo {volume} {7}},\ \bibinfo {pages} {1} (\bibinfo {year} {2021})}\BibitemShut {NoStop}%
\bibitem [{\citenamefont {Trotzky}\ \emph {et~al.}(2012)\citenamefont {Trotzky}, \citenamefont {Chen}, \citenamefont {Flesch}, \citenamefont {McCulloch}, \citenamefont {Schollwöck}, \citenamefont {Eisert},\ and\ \citenamefont {Bloch}}]{trotzky_probing_2012}%
  \BibitemOpen
  \bibfield  {author} {\bibinfo {author} {\bibfnamefont {S.}~\bibnamefont {Trotzky}}, \bibinfo {author} {\bibfnamefont {Y.-A.}\ \bibnamefont {Chen}}, \bibinfo {author} {\bibfnamefont {A.}~\bibnamefont {Flesch}}, \bibinfo {author} {\bibfnamefont {I.~P.}\ \bibnamefont {McCulloch}}, \bibinfo {author} {\bibfnamefont {U.}~\bibnamefont {Schollwöck}}, \bibinfo {author} {\bibfnamefont {J.}~\bibnamefont {Eisert}},\ and\ \bibinfo {author} {\bibfnamefont {I.}~\bibnamefont {Bloch}},\ }\bibfield  {title} {\bibinfo {title} {Probing the relaxation towards equilibrium in an isolated strongly correlated one-dimensional {Bose} gas},\ }\href {https://doi.org/10.1038/nphys2232} {\bibfield  {journal} {\bibinfo  {journal} {Nature Physics}\ }\textbf {\bibinfo {volume} {8}},\ \bibinfo {pages} {325} (\bibinfo {year} {2012})}\BibitemShut {NoStop}%
\bibitem [{noa()}]{noauthor_theoretical_nodate}%
  \BibitemOpen
  \href {https://www.theorie.physik.uni-muenchen.de/lsschollwoeck/} {{\selectlanguage {en}\bibinfo {title} {Theoretical {Nanophysics} - {LMU} {Munich}}}}\BibitemShut {NoStop}%
\bibitem [{\citenamefont {Vidal}(2003)}]{vidal_efficient_2003}%
  \BibitemOpen
  \bibfield  {author} {\bibinfo {author} {\bibfnamefont {G.}~\bibnamefont {Vidal}},\ }\bibfield  {title} {\bibinfo {title} {Efficient {Classical} {Simulation} of {Slightly} {Entangled} {Quantum} {Computations}},\ }\href {https://doi.org/10.1103/PhysRevLett.91.147902} {\bibfield  {journal} {\bibinfo  {journal} {Physical Review Letters}\ }\textbf {\bibinfo {volume} {91}},\ \bibinfo {pages} {147902} (\bibinfo {year} {2003})}\BibitemShut {NoStop}%
\bibitem [{\citenamefont {Vidal}(2004)}]{vidal_efficient_2004}%
  \BibitemOpen
  \bibfield  {author} {\bibinfo {author} {\bibfnamefont {G.}~\bibnamefont {Vidal}},\ }\bibfield  {title} {\bibinfo {title} {Efficient {Simulation} of {One}-{Dimensional} {Quantum} {Many}-{Body} {Systems}},\ }\href {https://doi.org/10.1103/PhysRevLett.93.040502} {\bibfield  {journal} {\bibinfo  {journal} {Physical Review Letters}\ }\textbf {\bibinfo {volume} {93}},\ \bibinfo {pages} {040502} (\bibinfo {year} {2004})}\BibitemShut {NoStop}%
\bibitem [{\citenamefont {Urbanek}\ and\ \citenamefont {Soldán}(2016)}]{urbanek_parallel_2016}%
  \BibitemOpen
  \bibfield  {author} {\bibinfo {author} {\bibfnamefont {M.}~\bibnamefont {Urbanek}}\ and\ \bibinfo {author} {\bibfnamefont {P.}~\bibnamefont {Soldán}},\ }\bibfield  {title} {\bibinfo {title} {Parallel implementation of the time-evolving block decimation algorithm for the {Bose}–{Hubbard} model},\ }\href {https://doi.org/10.1016/j.cpc.2015.10.016} {\bibfield  {journal} {\bibinfo  {journal} {Computer Physics Communications}\ }\textbf {\bibinfo {volume} {199}},\ \bibinfo {pages} {170} (\bibinfo {year} {2016})}\BibitemShut {NoStop}%
\bibitem [{\citenamefont {Sun}\ \emph {et~al.}()\citenamefont {Sun}, \citenamefont {Shirakawa},\ and\ \citenamefont {Yunoki}}]{sun_improved_2023}%
  \BibitemOpen
  \bibfield  {author} {\bibinfo {author} {\bibfnamefont {R.-Y.}\ \bibnamefont {Sun}}, \bibinfo {author} {\bibfnamefont {T.}~\bibnamefont {Shirakawa}},\ and\ \bibinfo {author} {\bibfnamefont {S.}~\bibnamefont {Yunoki}},\ }\href {https://doi.org/10.48550/arXiv.2312.02667} {\bibinfo {title} {Improved real-space parallelizable matrix-product state compression and its application to unitary quantum dynamics simulation}},\ \bibinfo {note} {arXiv:2312.02667}\BibitemShut {NoStop}%
\bibitem [{\citenamefont {Eisert}(2017)}]{eisert_towards_2017}%
  \BibitemOpen
  \bibfield  {author} {\bibinfo {author} {\bibfnamefont {J.}~\bibnamefont {Eisert}},\ }\href {https://www.youtube.com/watch?v=m0MOIgn11QA} {\bibinfo {title} {Towards quantum advantages of synthetic quantum systems}} (\bibinfo {year} {2017}),\ \bibinfo {note} {{IBM} {Research} {THINKQ} {Conference} on ``{Approximate} quantum computing: from advantage to applications''}\BibitemShut {NoStop}%
\bibitem [{\citenamefont {{University of Bath}}()}]{university_of_bath_balena_nodate}%
  \BibitemOpen
  \bibfield  {author} {\bibinfo {author} {\bibnamefont {{University of Bath}}},\ }\href {https://www.bath.ac.uk/corporate-information/balena-hpc-cluster/} {\bibinfo {title} {Balena {HPC} cluster}}\BibitemShut {NoStop}%
\bibitem [{\citenamefont {Secular}(2024)}]{secular_parallel_2024}%
  \BibitemOpen
  \bibfield  {author} {\bibinfo {author} {\bibfnamefont {P.}~\bibnamefont {Secular}},\ }\emph {\bibinfo {title} {Parallel tensor network methods for quantum lattice systems: matrix product state simulations on a supercomputer}},\ \href@noop {} {\bibinfo {type} {{PhD} thesis}},\ \bibinfo  {school} {University of Bath}, \bibinfo {address} {Bath} (\bibinfo {year} {2024})\BibitemShut {NoStop}%
\bibitem [{\citenamefont {Jaksch}\ \emph {et~al.}(1998)\citenamefont {Jaksch}, \citenamefont {Bruder}, \citenamefont {Cirac}, \citenamefont {Gardiner},\ and\ \citenamefont {Zoller}}]{jaksch_cold_1998}%
  \BibitemOpen
  \bibfield  {author} {\bibinfo {author} {\bibfnamefont {D.}~\bibnamefont {Jaksch}}, \bibinfo {author} {\bibfnamefont {C.}~\bibnamefont {Bruder}}, \bibinfo {author} {\bibfnamefont {J.~I.}\ \bibnamefont {Cirac}}, \bibinfo {author} {\bibfnamefont {C.~W.}\ \bibnamefont {Gardiner}},\ and\ \bibinfo {author} {\bibfnamefont {P.}~\bibnamefont {Zoller}},\ }\bibfield  {title} {\bibinfo {title} {Cold {Bosonic} {Atoms} in {Optical} {Lattices}},\ }\href {https://doi.org/10.1103/PhysRevLett.81.3108} {\bibfield  {journal} {\bibinfo  {journal} {Physical Review Letters}\ }\textbf {\bibinfo {volume} {81}},\ \bibinfo {pages} {3108} (\bibinfo {year} {1998})}\BibitemShut {NoStop}%
\bibitem [{\citenamefont {Schmitteckert}(2012)}]{schmitteckert_relaxation_2012}%
  \BibitemOpen
  \bibfield  {author} {\bibinfo {author} {\bibfnamefont {P.}~\bibnamefont {Schmitteckert}},\ }\bibfield  {title} {\bibinfo {title} {On the relaxation toward equilibrium in an isolated strongly correlated one-dimensional {Bose} gas},\ }\href {https://doi.org/10.1088/0031-8949/2012/T151/014059} {\bibfield  {journal} {\bibinfo  {journal} {Physica Scripta}\ }\textbf {\bibinfo {volume} {T151}},\ \bibinfo {pages} {014059} (\bibinfo {year} {2012})}\BibitemShut {NoStop}%
\bibitem [{\citenamefont {Cramer}\ \emph {et~al.}(2008{\natexlab{a}})\citenamefont {Cramer}, \citenamefont {Dawson}, \citenamefont {Eisert},\ and\ \citenamefont {Osborne}}]{cramer_exact_2008}%
  \BibitemOpen
  \bibfield  {author} {\bibinfo {author} {\bibfnamefont {M.}~\bibnamefont {Cramer}}, \bibinfo {author} {\bibfnamefont {C.~M.}\ \bibnamefont {Dawson}}, \bibinfo {author} {\bibfnamefont {J.}~\bibnamefont {Eisert}},\ and\ \bibinfo {author} {\bibfnamefont {T.~J.}\ \bibnamefont {Osborne}},\ }\bibfield  {title} {\bibinfo {title} {Exact {Relaxation} in a {Class} of {Nonequilibrium} {Quantum} {Lattice} {Systems}},\ }\href {https://doi.org/10.1103/PhysRevLett.100.030602} {\bibfield  {journal} {\bibinfo  {journal} {Physical Review Letters}\ }\textbf {\bibinfo {volume} {100}},\ \bibinfo {pages} {030602} (\bibinfo {year} {2008}{\natexlab{a}})}\BibitemShut {NoStop}%
\bibitem [{\citenamefont {Cramer}\ \emph {et~al.}(2008{\natexlab{b}})\citenamefont {Cramer}, \citenamefont {Flesch}, \citenamefont {McCulloch}, \citenamefont {Schollwöck},\ and\ \citenamefont {Eisert}}]{cramer_exploring_2008}%
  \BibitemOpen
  \bibfield  {author} {\bibinfo {author} {\bibfnamefont {M.}~\bibnamefont {Cramer}}, \bibinfo {author} {\bibfnamefont {A.}~\bibnamefont {Flesch}}, \bibinfo {author} {\bibfnamefont {I.~P.}\ \bibnamefont {McCulloch}}, \bibinfo {author} {\bibfnamefont {U.}~\bibnamefont {Schollwöck}},\ and\ \bibinfo {author} {\bibfnamefont {J.}~\bibnamefont {Eisert}},\ }\bibfield  {title} {\bibinfo {title} {Exploring {Local} {Quantum} {Many}-{Body} {Relaxation} by {Atoms} in {Optical} {Superlattices}},\ }\href {https://doi.org/10.1103/PhysRevLett.101.063001} {\bibfield  {journal} {\bibinfo  {journal} {Physical Review Letters}\ }\textbf {\bibinfo {volume} {101}},\ \bibinfo {pages} {063001} (\bibinfo {year} {2008}{\natexlab{b}})}\BibitemShut {NoStop}%
\bibitem [{\citenamefont {Flesch}\ \emph {et~al.}(2008)\citenamefont {Flesch}, \citenamefont {Cramer}, \citenamefont {McCulloch}, \citenamefont {Schollwöck},\ and\ \citenamefont {Eisert}}]{flesch_probing_2008}%
  \BibitemOpen
  \bibfield  {author} {\bibinfo {author} {\bibfnamefont {A.}~\bibnamefont {Flesch}}, \bibinfo {author} {\bibfnamefont {M.}~\bibnamefont {Cramer}}, \bibinfo {author} {\bibfnamefont {I.~P.}\ \bibnamefont {McCulloch}}, \bibinfo {author} {\bibfnamefont {U.}~\bibnamefont {Schollwöck}},\ and\ \bibinfo {author} {\bibfnamefont {J.}~\bibnamefont {Eisert}},\ }\bibfield  {title} {\bibinfo {title} {Probing local relaxation of cold atoms in optical superlattices},\ }\href {https://doi.org/10.1103/PhysRevA.78.033608} {\bibfield  {journal} {\bibinfo  {journal} {Physical Review A}\ }\textbf {\bibinfo {volume} {78}},\ \bibinfo {pages} {033608} (\bibinfo {year} {2008})}\BibitemShut {NoStop}%
\bibitem [{\citenamefont {Moussa}\ \emph {et~al.}(2017)\citenamefont {Moussa}, \citenamefont {Sarovar}, \citenamefont {Freeman}, \citenamefont {Lu},\ and\ \citenamefont {Luhman}}]{moussa_realizing_2017}%
  \BibitemOpen
  \bibfield  {author} {\bibinfo {author} {\bibfnamefont {J.~E.}\ \bibnamefont {Moussa}}, \bibinfo {author} {\bibfnamefont {M.}~\bibnamefont {Sarovar}}, \bibinfo {author} {\bibfnamefont {C.~D.}\ \bibnamefont {Freeman}}, \bibinfo {author} {\bibfnamefont {T.-M.}\ \bibnamefont {Lu}},\ and\ \bibinfo {author} {\bibfnamefont {D.~R.}\ \bibnamefont {Luhman}},\ }\href {https://doi.org/10.2172/1733291} {{\selectlanguage {en}\emph {\bibinfo {title} {Realizing the {Power} of {Near}-{Term} {Quantum} {Technologies}}}}},\ \bibinfo {type} {Tech. Rep.}\ \bibinfo {number} {SAND-2017-12971}\ (\bibinfo  {institution} {Sandia National Lab. (SNL-NM), Albuquerque, NM; Sandia National Lab. (SNL-CA), Livermore, CA; Univ. of California, Berkeley, CA},\ \bibinfo {year} {2017})\BibitemShut {NoStop}%
\bibitem [{\citenamefont {Dabelow}\ and\ \citenamefont {Reimann}(2020)}]{dabelow_relaxation_2020}%
  \BibitemOpen
  \bibfield  {author} {\bibinfo {author} {\bibfnamefont {L.}~\bibnamefont {Dabelow}}\ and\ \bibinfo {author} {\bibfnamefont {P.}~\bibnamefont {Reimann}},\ }\bibfield  {title} {\bibinfo {title} {Relaxation {Theory} for {Perturbed} {Many}-{Body} {Quantum} {Systems} versus {Numerics} and {Experiment}},\ }\href {https://doi.org/10.1103/PhysRevLett.124.120602} {\bibfield  {journal} {\bibinfo  {journal} {Physical Review Letters}\ }\textbf {\bibinfo {volume} {124}},\ \bibinfo {pages} {120602} (\bibinfo {year} {2020})}\BibitemShut {NoStop}%
\bibitem [{\citenamefont {Secular}()}]{secular_comment_2020}%
  \BibitemOpen
  \bibfield  {author} {\bibinfo {author} {\bibfnamefont {P.}~\bibnamefont {Secular}},\ }\href {https://doi.org/10.48550/arXiv.2005.02681} {\bibinfo {title} {Comment on ``{Relaxation} theory for perturbed many-body quantum systems versus numerics and experiment''}},\ \bibinfo {note} {arXiv:2005.02681}\BibitemShut {NoStop}%
\bibitem [{\citenamefont {Trotzky}\ \emph {et~al.}()\citenamefont {Trotzky}, \citenamefont {Chen}, \citenamefont {Flesch}, \citenamefont {McCulloch}, \citenamefont {Schollwöck}, \citenamefont {Eisert},\ and\ \citenamefont {Bloch}}]{trotzky_probing_2011}%
  \BibitemOpen
  \bibfield  {author} {\bibinfo {author} {\bibfnamefont {S.}~\bibnamefont {Trotzky}}, \bibinfo {author} {\bibfnamefont {Y.-A.}\ \bibnamefont {Chen}}, \bibinfo {author} {\bibfnamefont {A.}~\bibnamefont {Flesch}}, \bibinfo {author} {\bibfnamefont {I.~P.}\ \bibnamefont {McCulloch}}, \bibinfo {author} {\bibfnamefont {U.}~\bibnamefont {Schollwöck}}, \bibinfo {author} {\bibfnamefont {J.}~\bibnamefont {Eisert}},\ and\ \bibinfo {author} {\bibfnamefont {I.}~\bibnamefont {Bloch}},\ }\href {https://doi.org/10.48550/arXiv.1101.2659} {\bibinfo {title} {Probing the relaxation towards equilibrium in an isolated strongly correlated {1D} {Bose} gas}},\ \bibinfo {note} {arXiv:1101.2659}\BibitemShut {NoStop}%
\bibitem [{\citenamefont {Daley}\ \emph {et~al.}(2004)\citenamefont {Daley}, \citenamefont {Kollath}, \citenamefont {Schollwöck},\ and\ \citenamefont {Vidal}}]{daley_time-dependent_2004}%
  \BibitemOpen
  \bibfield  {author} {\bibinfo {author} {\bibfnamefont {A.~J.}\ \bibnamefont {Daley}}, \bibinfo {author} {\bibfnamefont {C.}~\bibnamefont {Kollath}}, \bibinfo {author} {\bibfnamefont {U.}~\bibnamefont {Schollwöck}},\ and\ \bibinfo {author} {\bibfnamefont {G.}~\bibnamefont {Vidal}},\ }\bibfield  {title} {\bibinfo {title} {Time-dependent density-matrix renormalization-group using adaptive effective {Hilbert} spaces},\ }\bibfield  {journal} {\bibinfo  {journal} {Journal of Statistical Mechanics: Theory and Experiment}\ }\textbf {\bibinfo {volume} {P04005}},\ \href {https://doi.org/10.1088/1742-5468/2004/04/P04005} {10.1088/1742-5468/2004/04/P04005} (\bibinfo {year} {2004})\BibitemShut {NoStop}%
\bibitem [{\citenamefont {White}\ and\ \citenamefont {Feiguin}(2004)}]{white_real-time_2004}%
  \BibitemOpen
  \bibfield  {author} {\bibinfo {author} {\bibfnamefont {S.~R.}\ \bibnamefont {White}}\ and\ \bibinfo {author} {\bibfnamefont {A.~E.}\ \bibnamefont {Feiguin}},\ }\bibfield  {title} {\bibinfo {title} {Real-{Time} {Evolution} {Using} the {Density} {Matrix} {Renormalization} {Group}},\ }\href {https://doi.org/10.1103/PhysRevLett.93.076401} {\bibfield  {journal} {\bibinfo  {journal} {Physical Review Letters}\ }\textbf {\bibinfo {volume} {93}},\ \bibinfo {pages} {076401} (\bibinfo {year} {2004})}\BibitemShut {NoStop}%
\bibitem [{\citenamefont {Singh}\ \emph {et~al.}(2011)\citenamefont {Singh}, \citenamefont {Pfeifer},\ and\ \citenamefont {Vidal}}]{singh_tensor_2011}%
  \BibitemOpen
  \bibfield  {author} {\bibinfo {author} {\bibfnamefont {S.}~\bibnamefont {Singh}}, \bibinfo {author} {\bibfnamefont {R.~N.~C.}\ \bibnamefont {Pfeifer}},\ and\ \bibinfo {author} {\bibfnamefont {G.}~\bibnamefont {Vidal}},\ }\bibfield  {title} {\bibinfo {title} {Tensor network states and algorithms in the presence of a global {U}(1) symmetry},\ }\href {https://doi.org/10.1103/PhysRevB.83.115125} {\bibfield  {journal} {\bibinfo  {journal} {Physical Review B}\ }\textbf {\bibinfo {volume} {83}},\ \bibinfo {pages} {115125} (\bibinfo {year} {2011})}\BibitemShut {NoStop}%
\bibitem [{\citenamefont {Stoudenmire}\ and\ \citenamefont {White}(2010)}]{stoudenmire_minimally_2010}%
  \BibitemOpen
  \bibfield  {author} {\bibinfo {author} {\bibfnamefont {E.~M.}\ \bibnamefont {Stoudenmire}}\ and\ \bibinfo {author} {\bibfnamefont {S.~R.}\ \bibnamefont {White}},\ }\bibfield  {title} {\bibinfo {title} {Minimally entangled typical thermal state algorithms},\ }\href {https://doi.org/10.1088/1367-2630/12/5/055026} {\bibfield  {journal} {\bibinfo  {journal} {New Journal of Physics}\ }\textbf {\bibinfo {volume} {12}},\ \bibinfo {pages} {055026} (\bibinfo {year} {2010})}\BibitemShut {NoStop}%
\bibitem [{\citenamefont {Hauschild}\ and\ \citenamefont {Pollmann}(2018)}]{hauschild_efficient_2018}%
  \BibitemOpen
  \bibfield  {author} {\bibinfo {author} {\bibfnamefont {J.}~\bibnamefont {Hauschild}}\ and\ \bibinfo {author} {\bibfnamefont {F.}~\bibnamefont {Pollmann}},\ }\bibfield  {title} {{\selectlanguage {en}\bibinfo {title} {Efficient numerical simulations with {Tensor} {Networks}: {Tensor} {Network} {Python} ({TeNPy})}},\ }\href {https://doi.org/10.21468/SciPostPhysLectNotes.5} {\bibfield  {journal} {\bibinfo  {journal} {SciPost Physics Lecture Notes}\ ,\ \bibinfo {pages} {5}} (\bibinfo {year} {2018})}\BibitemShut {NoStop}%
\bibitem [{\citenamefont {{TeNPy Developers}}()}]{tenpy_developers_tensor_nodate}%
  \BibitemOpen
  \bibfield  {author} {\bibinfo {author} {\bibnamefont {{TeNPy Developers}}},\ }\href {https://github.com/tenpy/tenpy} {\bibinfo {title} {Tensor {Network} {Python} {Package} ({TeNPy})}}\BibitemShut {NoStop}%
\bibitem [{\citenamefont {Tamascelli}\ \emph {et~al.}(2015)\citenamefont {Tamascelli}, \citenamefont {Rosenbach},\ and\ \citenamefont {Plenio}}]{tamascelli_improved_2015}%
  \BibitemOpen
  \bibfield  {author} {\bibinfo {author} {\bibfnamefont {D.}~\bibnamefont {Tamascelli}}, \bibinfo {author} {\bibfnamefont {R.}~\bibnamefont {Rosenbach}},\ and\ \bibinfo {author} {\bibfnamefont {M.~B.}\ \bibnamefont {Plenio}},\ }\bibfield  {title} {\bibinfo {title} {Improved scaling of time-evolving block-decimation algorithm through reduced-rank randomized singular value decomposition},\ }\href {https://doi.org/10.1103/PhysRevE.91.063306} {\bibfield  {journal} {\bibinfo  {journal} {Physical Review E}\ }\textbf {\bibinfo {volume} {91}},\ \bibinfo {pages} {063306} (\bibinfo {year} {2015})}\BibitemShut {NoStop}%
\bibitem [{\citenamefont {Unfried}\ \emph {et~al.}(2023)\citenamefont {Unfried}, \citenamefont {Hauschild},\ and\ \citenamefont {Pollmann}}]{unfried_fast_2023}%
  \BibitemOpen
  \bibfield  {author} {\bibinfo {author} {\bibfnamefont {J.}~\bibnamefont {Unfried}}, \bibinfo {author} {\bibfnamefont {J.}~\bibnamefont {Hauschild}},\ and\ \bibinfo {author} {\bibfnamefont {F.}~\bibnamefont {Pollmann}},\ }\bibfield  {title} {\bibinfo {title} {Fast time evolution of matrix product states using the {QR} decomposition},\ }\href {https://doi.org/10.1103/PhysRevB.107.155133} {\bibfield  {journal} {\bibinfo  {journal} {Physical Review B}\ }\textbf {\bibinfo {volume} {107}},\ \bibinfo {pages} {155133} (\bibinfo {year} {2023})}\BibitemShut {NoStop}%
\bibitem [{\citenamefont {Ganahl}\ \emph {et~al.}(2023)\citenamefont {Ganahl}, \citenamefont {Beall}, \citenamefont {Hauru}, \citenamefont {Lewis}, \citenamefont {Wojno}, \citenamefont {Yoo}, \citenamefont {Zou},\ and\ \citenamefont {Vidal}}]{ganahl_density_2023}%
  \BibitemOpen
  \bibfield  {author} {\bibinfo {author} {\bibfnamefont {M.}~\bibnamefont {Ganahl}}, \bibinfo {author} {\bibfnamefont {J.}~\bibnamefont {Beall}}, \bibinfo {author} {\bibfnamefont {M.}~\bibnamefont {Hauru}}, \bibinfo {author} {\bibfnamefont {A.~G.}\ \bibnamefont {Lewis}}, \bibinfo {author} {\bibfnamefont {T.}~\bibnamefont {Wojno}}, \bibinfo {author} {\bibfnamefont {J.~H.}\ \bibnamefont {Yoo}}, \bibinfo {author} {\bibfnamefont {Y.}~\bibnamefont {Zou}},\ and\ \bibinfo {author} {\bibfnamefont {G.}~\bibnamefont {Vidal}},\ }\bibfield  {title} {\bibinfo {title} {Density {Matrix} {Renormalization} {Group} with {Tensor} {Processing} {Units}},\ }\href {https://doi.org/10.1103/PRXQuantum.4.010317} {\bibfield  {journal} {\bibinfo  {journal} {PRX Quantum}\ }\textbf {\bibinfo {volume} {4}},\ \bibinfo {pages} {010317} (\bibinfo {year} {2023})}\BibitemShut {NoStop}%
\bibitem [{\citenamefont {Stoudenmire}\ and\ \citenamefont {White}(2013)}]{stoudenmire_real-space_2013}%
  \BibitemOpen
  \bibfield  {author} {\bibinfo {author} {\bibfnamefont {E.~M.}\ \bibnamefont {Stoudenmire}}\ and\ \bibinfo {author} {\bibfnamefont {S.~R.}\ \bibnamefont {White}},\ }\bibfield  {title} {\bibinfo {title} {Real-space parallel density matrix renormalization group},\ }\href {https://doi.org/10.1103/PhysRevB.87.155137} {\bibfield  {journal} {\bibinfo  {journal} {Physical Review B}\ }\textbf {\bibinfo {volume} {87}},\ \bibinfo {pages} {155137} (\bibinfo {year} {2013})}\BibitemShut {NoStop}%
\bibitem [{\citenamefont {Depenbrock}(2013)}]{depenbrock_tensor_2013}%
  \BibitemOpen
  \bibfield  {author} {\bibinfo {author} {\bibfnamefont {S.}~\bibnamefont {Depenbrock}},\ }{\selectlanguage {de}\emph {\bibinfo {title} {Tensor networks for the simulation of strongly correlated systems}}},\ \href {https://edoc.ub.uni-muenchen.de/15963/} {\bibinfo {type} {{PhD} thesis}},\ \bibinfo  {school} {Ludwig-Maximilians-Universität München} (\bibinfo {year} {2013})\BibitemShut {NoStop}%
\bibitem [{\citenamefont {Secular}\ \emph {et~al.}(2020)\citenamefont {Secular}, \citenamefont {Gourianov}, \citenamefont {Lubasch}, \citenamefont {Dolgov}, \citenamefont {Clark},\ and\ \citenamefont {Jaksch}}]{secular_parallel_2020}%
  \BibitemOpen
  \bibfield  {author} {\bibinfo {author} {\bibfnamefont {P.}~\bibnamefont {Secular}}, \bibinfo {author} {\bibfnamefont {N.}~\bibnamefont {Gourianov}}, \bibinfo {author} {\bibfnamefont {M.}~\bibnamefont {Lubasch}}, \bibinfo {author} {\bibfnamefont {S.}~\bibnamefont {Dolgov}}, \bibinfo {author} {\bibfnamefont {S.~R.}\ \bibnamefont {Clark}},\ and\ \bibinfo {author} {\bibfnamefont {D.}~\bibnamefont {Jaksch}},\ }\bibfield  {title} {\bibinfo {title} {Parallel time-dependent variational principle algorithm for matrix product states},\ }\href {https://doi.org/10.1103/PhysRevB.101.235123} {\bibfield  {journal} {\bibinfo  {journal} {Physical Review B}\ }\textbf {\bibinfo {volume} {101}},\ \bibinfo {pages} {235123} (\bibinfo {year} {2020})}\BibitemShut {NoStop}%
\bibitem [{\citenamefont {Goto}\ and\ \citenamefont {Danshita}(2019)}]{goto_performance_2019}%
  \BibitemOpen
  \bibfield  {author} {\bibinfo {author} {\bibfnamefont {S.}~\bibnamefont {Goto}}\ and\ \bibinfo {author} {\bibfnamefont {I.}~\bibnamefont {Danshita}},\ }\bibfield  {title} {\bibinfo {title} {Performance of the time-dependent variational principle for matrix product states in the long-time evolution of a pure state},\ }\href {https://doi.org/10.1103/PhysRevB.99.054307} {\bibfield  {journal} {\bibinfo  {journal} {Physical Review B}\ }\textbf {\bibinfo {volume} {99}},\ \bibinfo {pages} {054307} (\bibinfo {year} {2019})}\BibitemShut {NoStop}%
\bibitem [{noa(2018)}]{noauthor_university_2018}%
  \BibitemOpen
  \href {https://doi.org/10.15125/b6cd-s854} {\bibinfo {title} {University of {Bath}, {Research} {Computing}}} (\bibinfo {year} {2018})\BibitemShut {NoStop}%
\bibitem [{\citenamefont {Al-Assam}\ \emph {et~al.}(2017)\citenamefont {Al-Assam}, \citenamefont {Clark},\ and\ \citenamefont {Jaksch}}]{al-assam_tensor_2017}%
  \BibitemOpen
  \bibfield  {author} {\bibinfo {author} {\bibfnamefont {S.}~\bibnamefont {Al-Assam}}, \bibinfo {author} {\bibfnamefont {S.~R.}\ \bibnamefont {Clark}},\ and\ \bibinfo {author} {\bibfnamefont {D.}~\bibnamefont {Jaksch}},\ }\bibfield  {title} {\bibinfo {title} {The tensor network theory library},\ }\bibfield  {journal} {\bibinfo  {journal} {Journal of Statistical Mechanics: Theory and Experiment}\ }\textbf {\bibinfo {volume} {093102}},\ \href {https://doi.org/10.1088/1742-5468/aa7df3} {10.1088/1742-5468/aa7df3} (\bibinfo {year} {2017})\BibitemShut {NoStop}%
\bibitem [{\citenamefont {Goodyer}(2013)}]{goodyer_tnt_2013}%
  \BibitemOpen
  \bibfield  {author} {\bibinfo {author} {\bibfnamefont {C.}~\bibnamefont {Goodyer}},\ }\href {http://www.hector.ac.uk/cse/distributedcse/reports/UniTNT/} {\emph {\bibinfo {title} {{TNT} {Library}: {Tensor} {Manipulation} and {Storage}}}},\ \bibinfo {type} {Tech. Rep.}\ (\bibinfo  {institution} {NAG Ltd},\ \bibinfo {year} {2013})\BibitemShut {NoStop}%
\bibitem [{\citenamefont {Gobert}\ \emph {et~al.}(2005)\citenamefont {Gobert}, \citenamefont {Kollath}, \citenamefont {Schollwöck},\ and\ \citenamefont {Schütz}}]{gobert_real-time_2005}%
  \BibitemOpen
  \bibfield  {author} {\bibinfo {author} {\bibfnamefont {D.}~\bibnamefont {Gobert}}, \bibinfo {author} {\bibfnamefont {C.}~\bibnamefont {Kollath}}, \bibinfo {author} {\bibfnamefont {U.}~\bibnamefont {Schollwöck}},\ and\ \bibinfo {author} {\bibfnamefont {G.}~\bibnamefont {Schütz}},\ }\bibfield  {title} {\bibinfo {title} {Real-time dynamics in spin-{$\frac{1}{2}$} chains with adaptive time-dependent density matrix renormalization group},\ }\href {https://doi.org/10.1103/PhysRevE.71.036102} {\bibfield  {journal} {\bibinfo  {journal} {Physical Review E}\ }\textbf {\bibinfo {volume} {71}},\ \bibinfo {pages} {036102} (\bibinfo {year} {2005})}\BibitemShut {NoStop}%
\end{thebibliography}%

\end{document}